\documentclass[letterpaper,11pt]{article}
\usepackage{caption}
\pdfoutput=1 % if your are submitting a pdflatex (i.e. if you have
\usepackage{extarrows}
\usepackage{amssymb}
\usepackage{amsmath}
\usepackage{setspace}

\usepackage{amstext}
\usepackage{etoolbox}
\let\tinymatrix\smallmatrix

\patchcmd{\tinymatrix}{\scriptstyle}{\scriptscriptstyle}{}{}
\patchcmd{\tinymatrix}{\scriptstyle}{\scriptscriptstyle}{}{}
\patchcmd{\tinymatrix}{\vcenter}{\vtop}{}{}
\patchcmd{\tinymatrix}{\bgroup}{\bgroup\scriptsize}{}{}

\usepackage{graphicx,epsfig}
\usepackage{epsfig}
\usepackage{verbatim} 
\usepackage{color}
\usepackage{ulem}
\usepackage{enumitem}
\usepackage{subfigure}
\usepackage{varwidth,xcolor}
\usepackage{bbm}
\usepackage{parskip}
\usepackage{dsfont}
\usepackage[numbers,sort&compress]{natbib}
\usepackage[body={6.5in, 9in}, right=1in, top=1in]{geometry}
\usepackage{mathtools}
\usepackage{comment}
\usepackage{caption}
\usepackage{bbold}
\usepackage{float}

\newcommand{\Comment}[1]{{}}
\definecolor{darkblue}{rgb}{0.15,0.35,0.55}
\definecolor{darkgreen}{rgb}{0.2,0.7,0.3}
\definecolor{reddish}{rgb}{0.65, 0.2, 0.2}
\usepackage[linktocpage=true]{hyperref}
\newcommand{\be}{\begin{equation}}
\newcommand{\ee}{\end{equation}}
\newcommand{\bea}{\begin{eqnarray}}
\newcommand{\eea}{\end{eqnarray}}
\newcommand{\beas}{\begin{eqnarray*}}
\newcommand{\eeas}{\end{eqnarray*}}

\def\({\left(}
\def\){\right)}

\newcommand{\rd}{{\rm d}}

\newcommand\bigzero{\makebox(0,0){\text{\huge0}}}

\def\gsim{ \lower .75ex \hbox{$\sim$} \llap{\raise .27ex \hbox{$>$}} }
\def\lsim{ \lower .75ex \hbox{$\sim$} \llap{\raise .27ex \hbox{$<$}} }

\usepackage{xcolor,colortbl}

\begin{document}
\def\thefootnote{\fnsymbol{footnote}}

\begin{center}
\LARGE{\textbf{Early-Time Measure in Eternal Inflation}} \\[0.5cm]
 
\large{Justin Khoury and Sam S. C. Wong}
\\[0.5cm]

\small{
\textit{Center for Particle Cosmology, Department of Physics and Astronomy, University of Pennsylvania,\\ Philadelphia, PA 19104}}

\vspace{.2cm}

\end{center}

\vspace{.6cm}

\hrule \vspace{0.2cm}
\centerline{\small{\bf Abstract}}
%\vspace{-0.2cm}
{\small\noindent In a situation like eternal inflation, where our data is replicated at infinitely-many other space-time events, it is necessary to make a prior assumption about our location to extract predictions. The principle of mediocrity entails that we live at asymptotic late times, when the occupational probabilities of vacua has settled to a near-equilibrium distribution. In this paper we further develop the idea that we instead exist during the approach to equilibrium, much earlier than the exponentially-long mixing time. In this case we are most likely to reside in vacua that are easily accessed dynamically. Using first-passage statistics, we prove that vacua that maximize their space-time volume at early times have: 1.~maximal ever-hitting probability; 2.~minimal mean first-passage time; and 3.~minimal decay rate. These requirements are succinctly captured by an early-time measure. The idea that we live at early times is a predictive guiding principle, with many phenomenological implications. First, our vacuum should lie deep in a funneled region, akin to folding energy landscapes of proteins. Second, optimal landscape regions are characterized by relatively short-lived vacua, with lifetime of order the de Sitter Page time. For our vacuum, this lifetime is~$\sim 10^{130}$~years, which is consistent with the Standard Model estimate due to Higgs metastability. Third, the measure favors vacua with small, positive vacuum energy. This can address the cosmological constant problem, provided there are sufficiently many vacua in the entire ensemble of funnels. As a concrete example, we study the Bousso-Polchinski lattice of flux vacua, and find that the early-time measure favors lattices with the fewest number of flux dimensions. This favors compactifications with a large hierarchy between the lightest modulus and all other K\"ahler and complex structure moduli.
\vspace{0.3cm}
\noindent
\hrule
\def\thefootnote{\arabic{footnote}}
\setcounter{footnote}{0}

\section{Introduction} 

A striking feature of our universe is its simplicity. The discovery of a 125~GeV Higgs boson together with the absence of new physics in precision and LHC data points to the Standard Model (SM) being valid up to very high energy. On the gravitational side, General Relativity (GR) has passed every observational test with flying colors, most recently in the strong field regime. The driver of cosmic acceleration appears to be a cosmological constant (CC), the simplest form of dark energy. Surprises may be lurking, but all the evidence so far points to a simple, minimal universe.

Another striking feature of our universe is its near-criticality. When extrapolated to high energy, the Higgs quartic coupling becomes negative at~$\sim 10^{10}$~GeV and remains small, which implies metastability~\cite{Frampton:1976kf,Sher:1988mj,Casas:1994qy,Espinosa:1995se,Isidori:2001bm,Espinosa:2007qp,Ellis:2009tp,Degrassi:2012ry,Buttazzo:2013uya,Lalak:2014qua,Andreassen:2014gha,Branchina:2014rva,Bednyakov:2015sca,Iacobellis:2016eof,Andreassen:2017rzq}. The estimated lifetime,~$\tau_{\rm SM} = 10^{526^{+409}_{-202}}~{\rm years}$~\cite{Andreassen:2017rzq}, while reassuringly long, hinges on an exquisitely delicate cancelation between an exponentially suppressed tunneling rate and the exponentially-large volume of the observable universe. Intriguingly, the SM lifetime relates the CC, which sets the observable 4-volume, and the weak scale, which sets the Higgs and top quark masses.   

It is remarkable that other fine-tunings can be interpreted as problems of near-criticality. The hierarchy problem translates to the Higgs having a nearly-vanishing mass (relative to~$M_{\rm Pl}$), close to the phase transition between broken/unbroken electroweak symmetry~\cite{Giudice:2006sn}. Less sharply defined, yet equally intuitive, is the CC problem, which translates to our universe being nearly flat, Minkowski space, which delineates between de Sitter (dS) and anti-de Sitter (AdS) space-times with different asymptotics and stability properties~\cite{Friedrich:1987,Bizon:2011gg}. Lastly, slow-roll inflation, long understood as a critical phenomenon~\cite{Strominger:2001pn}, interpolates between an approximate dS conformal fixed point and standard decelerating expansion.

The near-criticality of our universe strongly suggests a statistical physics approach. A natural arena for the statistical physics of universes
is the vast landscape of string theory, together with the mechanism of eternal inflation~\cite{Steinhardt:1982kg,Vilenkin:1983xq,Linde:1986fc,Linde:1986fd,Starobinsky:1986fx}
for dynamically populating these vacua. Much remains to be understood about the string landscape, in particular the subtle constraints that quantum gravity places on low-energy physics beyond the usual rules of effective field theory (EFT)~\cite{Obied:2018sgi,Agrawal:2018own}. But it seems unambiguous that the landscape is comprised of an exponentially-large number of metastable states with correspondingly rich slew of possible low-energy physics. How is the simplicity/minimality of the SM selected from this plethora of complex EFTs with rich particle spectra?

\subsection{Quasi-equilibrium eternal inflation}

A long-standing challenge in extracting predictions in the multiverse is the {\it measure problem} inherent to eternal inflation~\cite{Guth:2007ng,Freivogel:2011eg}. Since pocket universes of all types are generated infinitely-many times in an eternally-inflating universe, relative probabilities require a regularization prescription (or measure). While a rigorous resolution to the measure problem may ultimately require a complete understanding of quantum gravity, theorists have in the meantime pursued various semi-classical prescriptions. The hope is to identify a class of well-behaved measures with more or less universal predictions that hopefully approximate those of the ``true" measure. 

Underlying nearly all existing approaches to the measure problem is the assumption that the multiverse has existed for a sufficiently long time, much longer than the exponentially-long mixing time, such that vacuum statistics have settled to a quasi-stationary/equilibrium distribution~\cite{Linde:1993nz,Linde:1993xx,GarciaBellido:1993wn,Vilenkin:1994ua,Garriga:1997ef,Garriga:2005av}. This is motivated by the {\it principle of mediocrity}~\cite{Vilenkin:1994ua}, which roughly states that our pocket universe should be typical among all incarnations of habitable vacua in the multiverse. Since the vast majority of habitable vacua are generated in the asymptotic future in an eternally-inflating universe, the principle of mediocrity implies that our vacuum is selected according to quasi-equilibrium statistics.

While not completely free of ambiguities, the quasi-equilibrium framework leads to a fairly consistent and intuitive picture: the late-time distribution is dominated by the most stable dS vacuum anywhere on the landscape. This is the so-called {\it dominant vacuum}.\footnote{We assume that this vacuum is unique, for simplicity, though there might be a cluster of such dominant vacua with nearly degenerate lifetimes.}
Since the dominant vacuum is unlikely to be itself anthropically hospitable, habitable vacua are populated by exceedingly rare upward transitions, followed by a sequence of downward transitions. Such upward transitions occur on a doubly-exponentially long time scale, given by the dS recurrence time for the dominant vacuum,~$t_{\rm rec} \sim H_{\rm dom}^{-1} {\rm e}^{8\pi^2 M_{\rm Pl}^2/H_{\rm dom}^2}$. This raises the specter of other rare, but far more likely, fluctuations, in particular Boltzmann brains (BBs)~\cite{Albrecht:2002uz,Dyson:2002pf,Albrecht:2004ke,Page:2005ur,Page:2006dt}. Indeed, BB domination represents the greatest phenomenological threat to the viability of late-time, quasi-equilibrium measures. In some cases, it can be avoided if the landscape satisfies a number of non-trivial conditions~\cite{DeSimone:2008if,Bousso:2008hz}, {\it e.g.}, all vacua (habitable or not) that support BBs should have a larger decay rate than the fastest BB production rate. Whether such conditions are satisfied in the string landscape remains an open question.
  
More generally, as emphasized by Hartle and Srednicki~\cite{Hartle:2007zv,Srednicki:2009vb}, in a situation like eternal inflation where our data is replicated at (infinitely-many) other space-time locations, it is necessary to make a prior assumption about our location within the multiverse to extract physical predictions. This can be formalized in a Bayesian approach by specifying a ``xerographic distribution"~\cite{Hartle:2007zv,Srednicki:2009vb}. To be clear, even if the fundamental theory handed us a unique and unambiguous measure, a xerographic distribution would still be necessary to translate from third-person probabilities dictated by the measure, to first-person probabilities specifying what {\it we} are most likely to observe. The principle of mediocrity corresponds in this language to a uniform xerographic distribution. But it is not the unique choice. 

\subsection{The approach to equilibrium}

Instead of focusing on quasi-equilibrium distributions, an alternative approach developed in~\cite{Khoury:2019yoo,Khoury:2019ajl,Kartvelishvili:2020thd} supposes that we live during the {\it approach to equilibrium}, much earlier than the exponentially-long mixing time for the landscape. Our working assumption is that our vacuum is typical among all habitable vacua that are accessed well before the mixing time, corresponding to a xerographic distribution that cuts off at times much earlier than the relaxation time. While this is certainly a more restrictive notion of typicality than the principle of mediocrity, it is a logical possibility whose consequences are worth exploring.\footnote{Of course, some notion of typicality is necessary and implicit in scientific experiments~\cite{Garriga:2007wz,Srednicki:2009vb}. But typicality is defined with respect to a reference class of observers, and how broad this reference class should be is a matter of assumption.} Indeed, a key point of~\cite{Hartle:2007zv,Srednicki:2009vb} is that different xerographic distributions can compete in a Bayesian framework, by comparing posterior probabilities given our current data or making predictions for future measurements. 

In short, as emphasized in~\cite{Denef:2017cxt}, the relevant question in the approach to equilibrium is, What habitable vacua have the right properties to be easily accessed early on in the evolution? This translates to a search optimization problem~\cite{Khoury:2019yoo}: accessible vacua reside in optimal regions where local landscape dynamics are efficient. This idea was formalized with the definition of an accessibility measure~\cite{Khoury:2019ajl}, which is the landscape analogue of the closeness centrality index~\cite{closeness1,closeness2} in network science. As we will make precise in this paper, optimal regions of the landscape display non-equilibrium critical phenomena, in the sense that their vacuum dynamics are tuned at {\it dynamical criticality}. This hints at a deep connection between the near-criticality of our universe and non-equilibrium critical phenomena on the landscape.\footnote{In an interesting recent paper, Giudice {\it et al.}~\cite{Giudice:2021viw} proposed an alternative mechanism, dubbed `self-organized localization', to explain the near-criticality of our universe. Interestingly, their approach is complementary to ours, as they study quantum first-order (instead of classical, second-order non-equilibrium) phase transitions in stochastic inflation (instead of false-vacuum eternal inflation).}
    
An immediate advantage of working at early times is that it circumvents the issue of BBs, since by assumption the relevant dynamics take place well before the recurrence time of any vacuum supporting BB production~\cite{Srednicki:2009vb}. Relatedly, a convenient simplification with working at early times is that it allows a ``downward" approximation~\cite{SchwartzPerlov:2006hi,Olum:2007yk}.
Since ``upward" transitions are doubly-exponentially suppressed relative to downward transitions, and occur on a time scale of the dS recurrence time~\cite{Lee:1987qc},  
they can be safely neglected for low-energy vacua of interest. The downward approximation will simplify many of our results.

On the flip side, a drawback of working at early times is sensitivity to initial conditions, which seems at odds with the attractor property of inflation. In this sense our approach is similar to all local measures, {\it e.g.},~\cite{Garriga:2005av,Bousso:2006ev}, which probe a finite region of the eternally-inflating space-time. While insensitivity to initial conditions is certainly desirable, it is logically distinct and not necessary to formulate a cosmological measure. In any case, our strategy is to identify general criteria for vacua to be accessed early on, applicable to a broad range of initial conditions. Our approach does rely, however, on the multiverse starting out in a low-entropy state, such that high energy vacua are initially populated. 

\subsection{Funneled landscapes and dynamical criticality}

In this paper we develop various aspects of the early-time framework, with the goal of making predictions that are distinct from the standard approach to eternal inflation and largely independent of anthropic reasoning. For this purpose, we will make use of {\it first-passage statistics}~\cite{Redner},  in particular the mean first-passage time (MFPT), which are ideally-suited to study the approach to equilibrium. 

After reviewing the formalism of vacuum dynamics on the landscape and essential elements of first-passage statistics in Sec.~\ref{RW complex nets}, we derive in Sec.~\ref{inequality} a lower bound satisfied by the comoving volume of vacua, valid in the downward approximation. This volume inequality is a key result of the paper. The lower bound is maximized for vacua having: 

\begin{enumerate}

\item Maximal ever-hitting probability, {\it i.e.}, they are easily accessed;

\item Minimal MFPT, {\it i.e.}, they are accessed early on;

\item Minimal decay rate, {\it i.e.}, they are long-lived.

\end{enumerate}
To capture these three requirements, we will introduce a dimensionless figure-of-merit, given by~\eqref{measure dS} and~\eqref{measure AdS} for dS and
AdS vacua, respectively. This figure-of-merit defines an {\it early-time measure}. 

To proceed, we focus in Sec.~\ref{opt and cri} on a finite region of the landscape, consisting of a large number of inflating and terminal vacua. Furthermore, we envision that a large ensemble of such regions exists in the vastness of the landscape, each having on average the same number of vacua but differing in their graph topology and transition rates. The optimization problem is to identify network properties that satisfy the above three requirements, and thereby maximize volume at early times. 

An important lesson can be drawn from the requirement of maximal ever-hitting probability. In the downward approximation, we will show that the ever-hitting probability can be expressed as a sum over all paths that start from high energy vacua, weighted by the branching probability for each path. Thus low-lying vacua will be accessed provided there exists a sequence of downward transitions from high-energy vacua. This will be the case if the landscape region has the topography of a funnel~\cite{Khoury:2019yoo,Khoury:2019ajl,Kartvelishvili:2020thd}, akin to the {\it principle of minimal frustration} in protein folding~\cite{proteins1}. Naturally-occurring proteins fold efficiently because their free energy landscape is characterized by a smooth funnel near the native state. 

Another key lesson comes from the interplay of minimizing the MFPT and minimizing decay rates. By itself, minimizing the MFPT is trivial.
The shortest possible time to access a vacuum is of order a few e-folds. But this would require its parent vacua to have lifetimes of order a few e-folds,
which is suboptimal from the point of view of minimizing decay rates. To quantify this tension, we will work in terms of an average MFPT, known as
Kemeny's constant~\cite{kemeny}, defined as the MFPT suitably averaged over target vacua. We will show that the optimal choice is achieved when the average MFPT
to low-lying vacua scales logarithmically with the number of vacua,~$\sim \log N$, which signals {\it dynamical criticality}. This also describes a {\it computational phase transition}~\cite{compPT1,compPT2} in how Kemeny's constant scales with moduli-space dimensionality~$D \sim \log N$. Indeed, in generic (frustrated) regions,~${\cal T}$ is expected to grow rapidly with~$D$, consistent with the~\textsf{NP}-hard complexity class of the general problem~\cite{Denef:2006ad,Halverson:2018cio}. Optimal regions, however, correspond to special, polynomial-time instances of the general problem. (See also~\cite{Bao:2017thx}.)

As first shown~\cite{Khoury:2019yoo}, the critical case corresponds to dS vacua having an average lifetime given by the dS Page time:~$\tau_{\rm crit} \sim M_{\rm Pl}^2/H^3$.
If our vacuum is part of an optimal region, then, taking the observed CC as an input, this would predict an optimal lifetime for our vacuum of~$M_{\rm Pl}^2/H_0^3 \sim 10^{130}~{\rm years}$. Remarkably, this is in good agreement with the SM estimate~\cite{Andreassen:2017rzq}:~$\tau_{\rm SM} = 10^{526^{+409}_{-202}}~{\rm years}$. The impact of new physics at the instability scale~$\mu_\lambda\sim 10^{10}$~GeV, in particular whether better agreement with the predicted optimal lifetime can be achieved, will be discussed elsewhere~\cite{thomas}. 

A further phenomenological implication of the early-time measure is the possibility of a non-anthropic solution to the CC problem. For fixed MFPT and ever-hitting probability, vacua occupying the most volume have the longest lifetime. On average, the lifetime of dS vacua is expected to grow with decreasing~$H$, as is the case, {\it e.g.}, for the critical lifetime~$\tau_{\rm crit}\sim M_{\rm Pl}^2/H^3$. Meanwhile, AdS regions crunch in a Hubble time, hence~$\tau_{\rm AdS} \sim |H|^{-1}$. This favors vacua with the smallest positive CCs across the ensemble of funnel regions. Therefore, if there are sufficiently many vacua in the entire ensemble of funneled regions,~$N_{\rm tot} \gg 10^{122}$, this would account for the inferred value of~$\sim 10^{-122}\,M_{\rm Pl}^4$ for the CC.

\subsection{Bousso-Polchinski flux landscapes}

As a concrete toy example of a funneled region, we consider in Sec.~\ref{BP section} the Bousso-Polchinski (BP) flux lattice~\cite{Bousso:2000xa}, {\it i.e.}, a discretuum of vacua with potential energy growing quadratically with distance from the origin. Within this framework, we calculate various first-passage observables of interest, as a function of the number~$D$ of flux directions (or, equivalently, field-space dimensions). We will find that regular landscape lattices with fewer field-space dimensions have longer escape time (stability), smaller Kemeny's constant (shorter mixing time), higher ever-hitting probability (accessibility), and overall higher early-time measure. Thus, regions with effective small field-space dimensionality are accessed quickly, and the space-time volume they occupy grows at a faster rate.  

We then explore the consequences for the number of light moduli and the effective dimensionality of moduli space. The string theory landscape is of course fundamentally of high dimensionality, in particular
due to the vast number of K\"ahler and complex structure moduli. If potential barriers are high enough to suppress tunneling along certain directions, then this would have the effect of reducing the effective field space dimensionality in the low energy theory, as indicated by the analysis of the BP lattice. We will demonstrate this with two classes of simple scalar potentials. The upshot is that the early-time framework favors compactifications with a large hierarchy between the lightest modulus, controlling transitions between vacua, and all other K\"ahler and complex structure moduli. 

\vspace{0.5cm}
\noindent For completeness, we have collected in Appendix~\ref{first-pass app} various results of first-passage statistics, including derivations of quantities used in the main text as well as other
identities that may prove useful.

\section{Vacuum dynamics on the landscape}
\label{RW complex nets}

The landscape can be represented mathematically as a network~\cite{Carifio:2017nyb}, whose nodes correspond to the different dS, AdS and Minkowski vacua. We assume as usual that AdS vacua are terminal, acting as absorbing nodes. Minkowski vacua, also generally assumed terminal, will be ignored for simplicity. Network edges represent the relevant transitions between vacua. (By ``relevant", we mean transitions with non-negligible rates on the time scale of interest.) For concreteness, we assume these are governed by Coleman-De Luccia (CDL) instantons~\cite{Coleman:1977py,Callan:1977pt,Coleman:1980aw}.\footnote{See~\cite{deAlwis:2019dkc,Cespedes:2020xpn} for interesting recent works on the subject.} In what follows, lower-case indices~$i,j,\ldots$ and~$a,b,\ldots$ will denote dS and AdS vacua respectively, while capital indices~$I,J\ldots$ will refer collectively to all vacua.

Following the seminal papers by Garriga, Vilenkin and collaborators~\cite{Garriga:1997ef,Garriga:2005av}, cosmological evolution on the landscape is described by a
Markov process. More precisely, because of terminal vacua, this process is an absorbing Markov process. The asymmetry between dS and AdS vacua means that
detailed balance does not hold, hence the dynamics are intrinsically non-equilibrium. The Markov process is defined in terms of a global time variable~$t$, related
to proper time in vacuum~$I$ via a lapse function:
\be
\Delta \tau_I = {\cal N}_I \Delta t\,. 
\ee
Although~$t$ is treated as a continuous parameter, the master equations below rely on coarse-graining over a time~$\Delta t$. For dS vacua,~$\Delta t$
should be long enough to average over transitory evolution between epochs of vacuum energy domination. For AdS vacua, since AdS bubbles crunch in a
Hubble time, coarse-graining spans their entire evolution. Thus an AdS bubble nucleated at a given time crunches and dies within a time~$\Delta t$ later.

In this paper we adopt scale-factor (or e-folding) time,
\be
t = \ln a\,, 
\ee
corresponding to~${\cal N}_i = H_i$. This is a popular choice in the literature, as it avoids various paradoxes, in particular the `youngness paradox'~\cite{Guth:2007ng}.
Using different lines of reasoning, it has been argued in stochastic inflation that scale-factor time is the preferred time variable~\cite{Finelli:2008zg,Finelli:2010sh,Vennin:2015hra}. In any case, with this choice 
the physical volume of dS vacua is simply related to their comoving volume: 
\be
{\cal V}^{\rm phys}_i(t) = {\rm e}^{3t} {\cal V}_i(t)\,.
\ee
Although the scale factor is not monotonic within AdS bubbles, this is irrelevant as we are coarse-graining over their evolution. And because AdS bubbles crunch
within a Hubble time, the only contribution to the {\it physical} volume of AdS vacua is limited to bubbles nucleated within the last e-fold ($\Delta t \simeq 1$).

With these assumptions, the master equations governing the comoving volume of dS and AdS vacua are respectively given by~\cite{Garriga:1997ef,Garriga:2005av}
\begin{subequations}
\label{master all}
\bea
\label{master dS}
& \displaystyle  \frac{{\rm d}{\cal V}_i}{{\rm d}t}  = \sum_j M_{ij} {\cal V}_j \,; \\
\label{master AdS}
& \displaystyle \frac{{\rm d}{\cal V}_a}{{\rm d}t}  = \sum_i \kappa_{ai}{\cal V}_i \,.
\eea
\end{subequations}
The~${\rm dS}\rightarrow {\rm dS}$ transition matrix is
\be
M_{ij}\equiv \kappa_{ij} - \delta_{ij} \kappa_j\,,
\label{Mdef}
\ee
where~$\kappa_{ij}$ is the~$j \rightarrow i$ transition rate, and~$\kappa_i \equiv \sum_J \kappa_{Ji}$ is the total decay rate of vacuum~$i$. (The latter includes decay channels into dS as well as AdS vacua.) 

For dS vacua, the solution to~\eqref{master dS} is
\be
{\cal V}_i(t) = \sum_j G_{ij}(t) {\cal V}_j(0)\,, 
\label{Vt V0}
\ee
where~${\cal V}_j(0)$ is the initial volume. The Green's function~$G_{ij}(t)$, given by
\be
G_{ij}(t) \equiv \left({\rm e}^{Mt}\right)_{ij}\,,
\label{Green def}
\ee
satisfies~\eqref{master dS} with initial condition~$G_{ij}(0) = \delta_{ij}$.  

Meanwhile, for AdS vacua, the solution to~\eqref{master AdS} is formally given by~${\cal V}_a(t) = \sum_i \int_0^t {\rm d}t' \kappa_{ai}{\cal V}_i(t')$. However, since the contribution to physical AdS volume is limited to bubbles nucleated within the last e-fold of evolution, as mentioned earlier, the only physically relevant part of AdS comoving volume at time~$t$ is
\be
{\cal V}_a(t) \simeq \sum_i \kappa_{ai} {\cal V}_i(t)\,.
\label{volume AdS}
\ee
Thus AdS vacua maximize their volume by being connected to parent dS vacua which 1) themselves occupy large volume; and 2) have significant
decay rate into their AdS offsprings. Nevertheless, since~$\kappa_{ai}  \; \lsim\; 1$ (and is generically~$\ll 1$), it is clear that AdS vacua occupy less volume than dS vacua.

\subsection{Transition rates and downward approximation}

Explicitly, transition rates between dS vacua are given by
\be
\kappa_{ij} = \frac{A_{ij}}{w_j} \,;\qquad A_{ij} = A_{ji} =  \left(\Lambda^4 {\rm e}^{-S_{\rm bounce}}\right)_{ij}\,,
\label{kappa dSdS}
\ee
where~$S_{\rm bounce}$ is the Euclidean action of the bounce solution, and~$\Lambda^4$ is the fluctuation determinant. 
For concreteness we have in mind transitions mediated by CDL instantons, as mentioned earlier, however note that~\eqref{kappa dSdS} also applies to
Hawking-Moss~\cite{Hawking:1981fz} and Brown-Teitelboim~\cite{Brown:1988kg} transitions. The `weight'~$w_j$ of the parent vacuum is given by
\be
w_j = H_j^4 {\rm e}^{S_j}\,,
\ee
where the factor of~$H_j^4$ converts the rate density to a rate, and~$S_j =  \frac{8\pi^2M_{\rm Pl}^2}{H_j^2}$ is the dS entropy of the parent vacuum. We will not need the explicit form of the~${\rm dS}\rightarrow {\rm AdS/Minkowski}$ transition rate, but suffice to say that it takes a similar form to~\eqref{kappa dSdS}, except of course that the numerator is not symmetric.

Although~$M_{ij}$ is not symmetric, its eigenvalues are nevertheless real~\cite{Garriga:2005av}. This can be seen by defining an auxiliary matrix
\be
\Sigma = W^{-1/2} M \, W^{1/2}\,,
\label{Sig M}
\ee
with~$W \equiv {\rm diag}(w_1,w_2,\ldots)$. This matrix is clearly symmetric and hence has real eigenvalues. Since~\eqref{Sig M} is a similarity transformation,~$M$ and~$\Sigma$ have identical spectra. 
Let us further assume that~$M$ is irreducible, {\it i.e.}, there exists a sequence of transitions connecting any pair of inflating vacua. In this case it
can be shown using Perron-Frobenius' theorem that its largest eigenvalue is non-degenerate and negative, while all other eigenvalues are strictly smaller:
\be
0 > \lambda_1 > \lambda_2 \geq \ldots 
\label{lamba's M}
\ee
The largest eigenvalue,~$\lambda_1$, sets the relaxation time for the Markov process. It vanishes if and only if the decay rate into terminals vanishes for all dS vacua.

The eigenvectors of $\Sigma$, denoted by~$v^{(\ell)}$, form a complete and orthonormal basis of the subspace of dS vacua.
As a further consequence of Perron-Frobenius' theorem, the components of the dominant eigenvector can all be positive:
\be
v^{(1)}_i \geq 0\,.
\label{v1 > 0}
\ee
Meanwhile, the eigenvectors of~$M$ are simply related to those of~$\Sigma$ via~$v^{(\ell)}_M = W^{1/2}v^{(\ell)}$.
Thus the eigenvectors of~$M$ also form a complete basis, albeit not orthonormal.

Importantly, because~$A_{ij}$ is symmetric~\cite{Lee:1987qc}, transitions between dS vacua satisfy a condition of detailed balance:
\be
\frac{\kappa_{ji}}{\kappa_{ij}} = \frac{w_j}{w_i}\sim {\rm e}^{S_j-S_i}\,.
\label{detailed balance}
\ee
Thus upward tunneling is doubly-exponentially suppressed. Notice that~\eqref{detailed balance} depends only on the false and true vacuum potential energy --- it is insensitive to the potential barrier and does not rely on the thin-wall approximation. In our early-time approach, we are justified in neglecting upward transitions (which decrease entropy) relative to downward transitions. This defines the ``downward approximation"~\cite{SchwartzPerlov:2006hi,Olum:2007yk}, which we will use repeatedly below. In this approximation, the network becomes a directed, acyclic graph.

To appreciate the technical simplification offered by the downward approximation, consider the eigenspectrum of the transition matrix~$M$. By labeling dS vacua in order of increasing potential energy,~$0 < V_1 \leq V_2 \leq \ldots$, the transition matrix reduces an upper-triangular matrix to leading order in the downward approximation:
\be
M \simeq \begin{bmatrix}
-\kappa_1 & \kappa_{12} & \kappa_{13} &  \ldots  \\
 & -\kappa_2 & \kappa_{23} &  \ldots \\
  & & -\kappa_3  & \ldots \\
 &  \raisebox{3ex}[0ex][0ex]{\bigzero} &  & \ddots
\end{bmatrix}  \,.
\label{M original}
\ee
Thus, instead of having to diagonalize a large~$N\times N$ matrix, the eigenvalues in the downward approximation can be read off from the diagonal entries. They are simply given by the decay rates of individual vacua,~$\{\lambda_1,\lambda_2,\ldots \} \simeq {\rm sort}\{-\kappa_1,-\kappa_2,\ldots\}$, where ``sort" stands for ordering the~$\kappa$'s from smallest to largest. In particular,~$\lambda_1$ is set by the most stable vacuum,
\be
\lambda_1 \simeq -\kappa_{\rm min}\,.
\ee
In the downward approximation it is possible for one or more dS vacua to become stable, {\it i.e.},~$\kappa_i \simeq 0$. This occurs whenever 
vacua have only up-tunneling as allowed transitions. We will  instead be interested in regions of the landscape with efficient dynamics, such that each
dS vacuum in the region has at least one allowed downward transition. Indeed, we will see that this is necessary to maximize the comoving volume of vacua.

\subsection{First-passage statistics}

The statistics ideally-suited to study the approach to equilibrium are {\it first-passage statistics}. In particular, the mean first-passage time (MFPT) is a
popular measure of search efficiency~\cite{Redner}. Aside from its present application in eternal inflation~\cite{Khoury:2019yoo,Khoury:2019ajl,Kartvelishvili:2020thd},
first-passage statistics have also been used in cosmology to study stochastic inflation~\cite{Vennin:2015hra,Assadullahi:2016gkk,Vennin:2016wnk,Noorbala:2018zlv}. For brevity we 
will mainly focus on~${\rm dS}\rightarrow {\rm dS}$ first-passage statistics, since this is all we will need in the remainder of the paper, and refer the reader to~\cite{Khoury:2019ajl} for~${\rm dS}\rightarrow {\rm AdS}$ results.

Of central importance is the {\it first-passage density},~$F_{ij}(t)$,~$i\neq j$, defined as the 
probability density that a random walker starting from~$j$ visits~$i$ for the first time at time~$t$.
Its integral is the first-passage probability, {\it i.e.}, the probability that the walker has visited~$i$ by time~$t$:
\be
{\cal P}_{ij}(t) \equiv \int_0^t {\rm d}t' F_{ij}(t')\,.
\label{first passage prob}
\ee
This probability is not necessarily normalized, since the {\it ever-hitting probability}~${\cal P}_{ij}^\infty = \int_0^\infty {\rm d}t' F_{ij}(t')$ 
is less than unity. All first-passage statistics can be derived from~$F$. For instance, the MFPT is given by its first moment:
\be
\langle t_{ij}\rangle \equiv \frac{\int_0^\infty {\rm d}t' \,t' F_{ij}(t')}{{\cal P}_{ij}^\infty} = - \left.\frac{{\rm d}\ln \tilde{F}_{ij}(s)}{{\rm d}s}\right\vert_{s=0} \,,
\label{MFPT def}
\ee
where the Laplace transform~$\tilde{F}_{ij}(s) = \int_0^\infty {\rm d}t\, F_{ij}(t) {\rm e}^{-st}$ is the generating function. Thus~$\langle t_{ij}\rangle$ is the average time to
reach~$i$ starting from~$j$, averaged over all paths connecting the pair. In other words, this is the average time conditioned on hitting the target $i$ since the conditional probability density~$\frac{F_{ij}(t)}{{\cal P}_{ij}^\infty}$ is used. 

A well-known result of random walk theory is the relation between the Green's function~\eqref{Green def} and the first-passage density~\cite{Redner}:
\be
G_{ij}(t) = \int_0^t {\rm d}t'\, G_{ii}(t-t') F_{ij}(t') \,; \qquad i\neq j\,.
\label{renewal}
\ee
The meaning is clear: the occupational probability at time~$t$ is the probability of reaching~$i$ at any earlier time~$t'$ multiplied by the 
loop probability~$G_{ii}(t-t')$ for returning to~$i$ in the remaining time. This relation simplifies in the downward approximation.
Since the network becomes acyclic in this approximation, the loop probability reduces to the survival probability,~$G_{ii}(t-t') \simeq {\rm e}^{-\kappa_i (t-t')}$,
hence
\be
G_{ij}(t)  \simeq \int_0^t {\rm d}t' {\rm e}^{-\kappa_i (t-t')}F_{ij}(t')\,.
\label{Green1}
\ee

It is instructive to consider an alternative derivation of \eqref{Green1}, as a number of intermediate steps will be useful later on. 
The solution for~$F_{ij}$ follows by taking the Laplace transform of~\eqref{renewal}:
\be
\tilde{F}_{ij}(s) = \frac{(s-M)^{-1}_{ij}}{(s-M)^{-1}_{ii}}\,.
\label{F(s)}
\ee
Meanwhile, the Laplace transform of the Green's function factorizes as
\be
(s - M)^{-1}_{ij} = (s+\kappa_i)^{-1} \left(\mathds{1} - T(s)\right)^{-1}_{ij} \,,
\label{Green}
\ee
where 
\be
T_{ij}(s) = \frac{\kappa_{ij}}{s+ \kappa_j} \,.
\ee
The matrix~$\left(\mathds{1} - T(s)\right)^{-1}$ is known in Markov chains as the fundamental matrix.
In the downward approximation, we clearly have~$\left(\mathds{1} - T(s)\right)^{-1}_{ii} \simeq 1$, hence 
\be
\tilde{F}_{ij}(s)\simeq \left(\mathds{1} - T(s)\right)^{-1}_{ij}\,.
\label{Fs}
\ee
Combining with~\eqref{Green} gives~$(s - M)^{-1}_{ij} \simeq (s+\kappa_i)^{-1} \tilde{F}_{ij}(s)$, which is the 
Laplace transform of~\eqref{Green1}, as desired.

As a corollary, notice that the ever-hitting probability in the downward approximation is
\be
{\cal P}_{ij}^\infty = \tilde{F}_{ij}(0) \simeq  \left(\mathds{1} - T\right)^{-1}_{ij} \,,
\label{P reln T}
\ee
where~$T_{ij} = \frac{\kappa_{ij}}{\kappa_j}$ is the~$j\rightarrow i$ branching ratio. Similarly, the MFPT in the downward approximation can be expressed as
\be
\langle t_{ij}\rangle \simeq \sum_k T_{ik} \kappa_k \left(\mathds{1} - T\right)^{-1}_{kj}\,.
\ee

\subsection{Alternative derivation of the MFPT}

For completeness, we offer an alternative derivation of the MFPT~$\langle t_{kj}\rangle$ between dS vacua. This derivation relies on the fact that, since we do not care about what happens 
after hitting~$k$ for the first, we may as well treat~$k$ itself as terminal. (The MFPT in this case is also known as the ``trapping time".) The resulting modified landscape has one fewer dS vacuum.  The transition matrix between the remaining dS vacua, denoted by~$M^{(k)}$, is obtained by scratching off the~$k$-th column and row of the original~$M$.

The first-passage density to terminal vacua is~\cite{Khoury:2019ajl}:~$F_{aj}(t) = \sum_i \kappa_{ai}G_{ij}(t)$.
Applying this result to~$k$ gives
\be
F_{kj}(t) = \sum_{i\neq k} \kappa_{ki} G_{ij}^{(k)}(t) \,;\qquad  G_{ij}^{(k)}(t) \equiv \left({\rm e}^{M^{(k)}t}\right)_{ij} \,,
\label{Fkj v2}
\ee
with Laplace transform 
\be
F_{kj}(s) =  \sum_{i\neq k} \kappa_{ki} \left(\frac{1}{s - M^{(k)}}\right)_{ij}\,.
\label{F(s) altern}
\ee
In fact~\eqref{Fkj v2} is a general expression for all kind of vacua. We give a probabilistic derivation of~\eqref{Fkj v2} in Appendix~\ref{first-pass app}. 
It is straightforward to show that this is equivalent to~\eqref{F(s)}, using the identity
\be
G_{ij}(t) = G_{ij}^{(k)}(t) + \int_0^t {\rm d}t'\, F_{kj}(t')  G_{ik}(t-t')\,.
\ee
Equation~\eqref{F(s) altern} then implies an alternative expression for the MFPT
\be
\langle t_{kj}\rangle  = - \frac{\sum_{i\neq k} \kappa_{ki} \left(M^{(k)}\right)^{-2}_{ij}}{\sum_{\ell \neq k} \kappa_{k\ell} \left(M^{(k)}\right)^{-1}_{\ell j}}\,.
\ee
 Note that the denominator $-\sum_{\ell \neq k} \kappa_{k\ell} \left(M^{(k)}\right)^{-1}_{\ell j} $ is the ever-hitting probability ${\cal P}^{\infty}_{kj}$. In the case that ${\cal P}^{\infty}_{kj}=1$ (no leakage to the environment), we have 
\be
\langle t_{kj}\rangle  = -\sum_{i\neq k} \left(M^{(k)}\right)^{-1}_{ij}.
\ee
This is a practically useful expression. In Appendix \ref{first-pass app} we use this expression to show that multiple target search time can be expressed solely in terms of single target first passage time. 
 
\subsection{Escape time} 

For some of the applications below, it will be useful to think about the {\it escape time}, {\it i.e.}, the characteristic time to escape some region~$\Omega$ comprised of~$N$ dS vacua. This can occur through decay either into terminals or into `environmental' dS vacua surrounding the region. As shown in Appendix~\ref{escape time proof}, the escape time from vacuum~$j$ in the region is given by
\be
t^{\rm esc}_j = -\sum_{i\in\Omega} M^{-1}_{ij} \,.
\label{Tesc j}
\ee
Furthermore, one can define a mean escape time, or {\it mean residency time}, for the region:
\be
\langle t_{\rm esc}\rangle = \frac{1}{N} \sum_{j\in\Omega} t^{\rm esc}_j =  \frac{1}{N} \sum_{i,j\in\Omega} M^{-1}_{ij} \,.
\label{Tesc}
\ee

\section{Maximizing volume at early times}
\label{inequality}

The core idea of our framework is that we live early on in the evolution of the multiverse, well before the mixing time for the entire landscape. Since we do not know {\it a priori} the time of evaluation, our goal is to define a measure for maximizing volume at early times. This measure, like most measures, should favor vacua occupying the most volume in space-time. Our strategy is to derive inequalities for the volume of vacua. We consider dS and AdS vacua in turn. 

Our starting point is equation~\eqref{Green1} for the Green's function, valid in the downward approximation, from which it follows that
\be
G_{ij}(t)  \;\gsim\; {\rm e}^{-\kappa_i t} {\cal P}_{ij}(t) \,,
\label{lower bound 1}
\ee
where~${\cal P}_{ij}(t)$ is the first-passage probability~\eqref{first passage prob}.  Equivalently, thinking of~$F_{ij}(t)/{\cal P}_{ij}^\infty$ as the (normalized) probability density function of hitting times,~${\cal P}_{ij}(t)/{\cal P}_{ij}^\infty$ is the corresponding cumulative distribution function, {\it i.e.}, the probability that the hitting time is less than~$t$. Markov's inequality for non-negative random variables informs us that
\be
{\cal P}_{ij}(t) \geq {\cal P}_{ij}^\infty \left( 1 - \frac{\langle t_{ij}\rangle}{t}\right)\,.
\label{lower bound 2}
\ee
Combining with~\eqref{lower bound 1}, we obtain
\be
G_{ij}(t)  \;\gsim\; {\rm e}^{-\kappa_i t} {\cal P}_{ij}^\infty \left( 1 - \frac{\langle t_{ij}\rangle}{t}\right)\,.
\label{lower bound}
\ee
This lower bound is non-trivial for~$\langle t_{ij}\rangle \leq t$. Hence, to maximize volume, the MFPT should be smaller than~$t$, which is itself much smaller than the global mixing time. On the other hand, it clearly follows from~\eqref{Green1} that the Green's function is bounded from above by the ever-hitting probability:
\be
G_{ij}(t)  \;\lsim\;  {\cal P}_{ij}^\infty\,.
\label{upper bound}
\ee
Thus we obtain the master inequality,
\be
{\rm e}^{-\kappa_i t} {\cal P}_{ij}^\infty \left( 1 - \frac{\langle t_{ij}\rangle}{t}\right) \;\lsim\; G_{ij}(t) \;\lsim\; {\cal P}_{ij}^\infty\,.
\label{master inequality 0}
\ee

To take initial conditions into account, it is helpful to define an initial occupational probability as the initial fraction of total comoving volume in dS vacua:
\be
p_j \equiv \frac{{\cal V}_j(0)}{{\cal V}_{\rm tot}}\,; \qquad \sum_j p_j = 1\,.
\ee
Consistent with our assumption of low-entropy initial conditions, we envision that the vacua that are initially populated tend to be high peaks.
The initial occupational probability allows us to average over initial conditions. For instance, we can define an average ever-hitting probability to~$i$,
\be
{\cal P}_{i}^\infty \equiv \sum_j p_j{\cal P}_{ij}^\infty\,.
\label{avg P}
\ee
Similarly we define an average MFPT as
\be
\langle t_{i}\rangle \equiv \sum_j p_j {\cal P}_{ij}^\infty \langle t_{ij}\rangle \,.
\label{average t_i}
\ee
Finally, it follows from~\eqref{Vt V0} that~$\sum_j p_j G_{ij}(t)  = \frac{{\cal V}_i(t)}{{\cal V}_{\rm tot}}$. Averaging~\eqref{master inequality 0}
in this way, we obtain
\be
\boxed{{\rm e}^{-\kappa_i t}  \left( {\cal P}_{i}^\infty - \frac{\langle t_{i}\rangle}{t}\right) \;\lsim\;\frac{{\cal V}_i(t)}{{\cal V}_{\rm tot}} \;\lsim\; {\cal P}_{i}^\infty\,.}
\label{master inequality}
\ee
This is a key result of this paper. As a proxy for maximizing volume, we seek vacua that maximize the lower and upper bounds. Clearly such vacua must have:

\begin{enumerate}

\item Maximal (downward) ever-hitting probability~${\cal P}_{i}^\infty$ --- they are easily accessed. 

\item Minimal MFPT~$\langle t_{i}\rangle$ --- they are accessed early on.

\item Minimal decay rate~$\kappa_i$ --- they are long-lived.

\end{enumerate}
A convenient quantity that captures these three requirements is
\be
\boxed{f_i \sim \frac{{\cal P}_{i}^\infty}{\kappa_i \langle t_{i}\rangle}\,.}
\label{measure dS}
\ee
We take this dimensionless figure of merit as our definition of an early-time measure for dS vacua.

Turning to AdS vacua, recall from~\eqref{volume AdS} that their volume is simply related to the volume of dS vacua that decay into them. It is natural to define
an early-time measure for AdS vacua as
\be
\boxed{f_a \sim   \sum_i \kappa_{ai} f_i \sim \sum_i T_{ai}  \frac{{\cal P}_{i}^\infty}{\langle t_{i}\rangle}\,,}
\label{measure AdS}
\ee
where~$T_{ai}  = \frac{\kappa_{ai}}{\kappa_i}$ is the~$i\rightarrow a$ branching ratio.

\section{Optimization and Criticality} 
\label{opt and cri}

To proceed, it is useful to think about a finite region of the landscape comprised of a large number~$N \gg 1$ of dS vacua, as well as some large number of terminal vacua.
We allow leakage into the environment but, for simplicity, ignore volume influx. In other words, random walkers on this finite network can escape
the region but never re-enter.\footnote{It is straightforward to account for environmental volume intake by including a stochastic source term. As shown in~\cite{Kartvelishvili:2020thd}, upon changing variables the master equation for dS vacua can be mapped to set of coupled stochastic oscillators described by the well-known Ornstein-Uhlenbeck process~\cite{OU}.} This is justified by the separation of time scales: as we are interested in search efficiency, the region can be defined as a set of vacua with healthy transition rates between them, relative to a slowly-evolving environment. 

In the vastness of the landscape, we envision a large ensemble of such regions, each having on average the same number of vacua but differing in their graph topology and transition rates. Our task is to identify the optimal network properties that maximize volume at early times. Concretely, we will explore the consequences of the three requirements derived in the previous Section, and argue that simultaneously satisfying these selects regions of the landscape that are tuned at criticality.

\subsection{Funnel topography}
\label{funnel sec}

Consider first the requirement of maximal ever-hitting probability. Combining~\eqref{P reln T} and~\eqref{avg P}, we can express~${\cal P}_{i}^\infty$
in terms of the branching probability:
\be
{\cal P}_{i}^\infty \simeq \lim_{s\rightarrow 0} \sum_j  p_j \left(\mathds{1} - T(s)\right)^{-1}_{ij} \,.
\label{P reln T avg}
\ee
Since~$T(s) = \frac{\kappa_{ij}}{s+ \kappa_j}$ is non-negative and~$\sum_i T_{ij}(s) < 1$, it follows from Perron-Frobenius' theorem that its largest eigenvalue is real, non-degenerate, and~$< 1$. Hence the fundamental matrix can be expanded as a geometric series~$\left(\mathds{1} - T(s)\right)^{-1} =\mathds{1} + T(s) + T^2(s) + \ldots$ The~$m^{\rm th}$ term in the series,~$\left(T^m(s)\right)_{ij}$, represents a branching probability for sequences~$j \rightarrow \ell_1\rightarrow\ldots \rightarrow\ell_{m-1}\rightarrow i$, summed over the~$m-1$ intermediaries.
Thus the ever-hitting probability becomes
\be
{\cal P}_{i}^\infty \simeq \sum_j  p_j  \sum_{m = 0}^\infty \; \sum_{\ell_{1},\ldots,\ell_{m-1}} T_{i\ell_{m-1}} \cdots T_{\ell_1 j} = \sum_j  p_j \sum_{\substack{\text{paths} \\ j\rightarrow i}} {\rm e}^{-S_{j\rightarrow i}}\,, 
\ee
where the sum is over all~$j\rightarrow i$ paths, with the ``action" for each path given by~$S_{j\rightarrow i} \equiv - \sum_{\rm edges} \log T$.
Equivalently, 
\be
{\cal P}_{i}^\infty \simeq \sum_j p_j \sum_{\substack{\text{paths} \\ j\rightarrow i}} \eta_{ij}\,,
\ee
where~$\eta \equiv \prod_{\rm edges} T$ is the branching probability for each path connecting~$j$ to~$i$.

Since this result assumed the downward approximation, the only contributing paths are paths describing a sequence of downward transitions. Thus a given vacuum~$i$ will be endowed with non-negligible volume {provided that there exists a downward path from at least one initially-populated vacuum~$j$}. Regions that maximize the volume of bulk vacua must therefore have the topography of a valley, or funnel~\cite{Khoury:2019yoo,Khoury:2019ajl,Kartvelishvili:2020thd}. This is akin to the smooth folding funnels of energy landscapes of naturally-occurring proteins~\cite{proteins1}.

\subsection{Kemeny's constant}
\label{Kemeny constant}

Next we turn to minimizing the MFPT in conjunction with minimizing decay rates. By itself, minimizing the MFPT is a trivial exercise.
The shortest possible time needed to access a vacuum is the coarse-graining time interval~$\Delta t$, which for scale-factor time is of
order a few e-folds. But this would require its parent vacua to have lifetimes of order~$\Delta t$, which is suboptimal from the point of
view of minimizing decay rates. In this Section we will argue that the optimal choice is achieved with an average MFPT to low-lying vacua of~${\cal O}(\ln N)$.

\vspace{0.3 cm}
\noindent {\bf Ignoring terminals and leakage}
\vspace{0.2cm}

\noindent To make this argument precise, it is convenient to first consider an identical landscape region but without terminals and ignoring leakage into the environment.
That is, this proxy region has only dS vacua and is a closed system. In what follows we will use hats to denote variables for this proxy region.
As shown in Appendix~\ref{F ineq appen}, in the downward approximation the first-passage densities with and without terminals/leakage satisfy the inequality:
\be
F_{ij}(t) \leq \hat{F}_{ij}(t)\,.
\label{F ineq}
\ee
This makes sense intuitively, as the presence of terminals/leakage decreases the first-passage density between dS vacua.
In particular, the first-moment inequality implies
\be
\langle t_{i}\rangle \leq   \langle \hat{t}_{i}\rangle \,.
\ee
Hence, instead of minimizing~$\langle t_{i}\rangle$, we can minimize its dS-only counterpart.  

The virtue of considering the proxy dS-only region is that it allows us to consider an average MFPT known as Kemeny's constant~\cite{kemeny}. Kemeny's constant is only defined for ergodic Markov processes, not absorbing ones, though below we will propose a modified Kemeny's constant for absorbing chains. In the proxy dS-only closed region, the transition matrix~$\hat{M}_{ij}$ admits a zero-mode~$\hat{f}_i^\infty$, which sets the stationary probability distribution:~$\sum_j \hat{M}_{ij} \hat{f}_j^\infty = 0$. The stationary distribution is related to the dominant eigenvector~$\hat{v}^{(1)}$ of the symmetric matrix~$\hat{\Sigma} = \hat{W}^{-1/2} \hat{M} \, \hat{W}^{1/2}$ via
\be
\hat{f}_i^\infty = \hat{v}^{(1)\,2}_i \,.
\ee

Kemeny's constant is defined by averaging~$\langle \hat{t}_{ij}\rangle$ over target vacua weighted by the stationary distribution~\cite{kemeny}:\footnote{Kemeny's constant is usually defined without the prefactor of~$1/N$, which we have included for convenience.} 
\be
\hat{{\cal T}} \equiv \frac{1}{N} \sum_{i\neq j} \langle \hat{t}_{ij}\rangle \hat{f}_i^\infty\,.
\label{average MFPT}
\ee
It is well-known that, remarkably,~$\hat{{\cal T}}$~is independent of the starting node~$j$, and as such is insensitive to initial conditions.
Furthermore, it is straightforward to show that Kemeny's constant is given by a spectral sum over the non-zero eigenvalues of the
transition matrix: 
\be
\hat{{\cal T}} =  \frac{1}{N}\sum_{\ell = 2}^{N} \frac{1}{|\hat{\lambda}_\ell|} \simeq  \frac{1}{N} \sum_{i \neq {\rm min}} \frac{1}{\hat{\kappa}_i}\,,
\label{average MFPT 2}
\ee
where in the last step we have used the downward approximation. The sum excludes the lowest-lying dS vacuum in the region, as it coincides with the zero-mode 
in the downward approximation. Equation~\eqref{average MFPT 2} is intuitively clear --- it states that, without terminals and in the downward approximation, 
the average MFPT simply reduces to the mean lifetime of dS vacua.

Now, one expects that generic regions of the landscape will feature one or more metastable dS vacua whose only possible transitions involve up-tunneling.
Such vacua become absolutely stable in the strict downward approximation, but more precisely their decay rate is doubly-exponentially suppressed at
sub-leading order, resulting in a doubly-exponentially large~$\hat{{\cal T}}$. Such regions are characterized by frustration and glassy dynamics~\cite{glassylandscape}. In contrast, optimal regions are characterized by a funneled landscape, as argued above, such that every metastable vacuum has at least one decay channel to a lower-lying dS vacuum. The resulting~$\hat{{\cal T}}$ is finite in the strict downward approximation. We henceforth focus our attention to such funneled landscapes.

\vspace{0.3 cm}
\noindent {\bf Modified Kemeny's constant with terminals and leakage}
\vspace{0.2cm}

\noindent Going back to the original region, our task is to generalize Kemeny's constant~\eqref{average MFPT} to the situation where there are terminals and leakage to environment~($\lambda_1<0$). A naive generalization of~\eqref{average MFPT}, namely~$\frac{1}{N} \sum_{i\neq j} \langle t_{ij}\rangle v^{(1)\,2}_i$, is no longer a constant, {\it i.e.}, it depends on the initial node~$j$. One option is to simply {\it define}~${\cal T}$ in terms of a spectral sum analogous to~\eqref{average MFPT 2}:
\be
{\cal T} \equiv \frac{1}{N}\sum_{\ell = 2}^{N} \frac{1}{|\lambda_\ell|} \simeq \frac{1}{N}  \sum_{i \neq {\rm min}} \frac{1}{\kappa_i}\,,
\label{average MFPT 3}
\ee
where the last step again assumes the downward approximation. Here,~$\kappa_i$ includes decay channels into terminal vacua, and the sum excludes the longest-lived vacuum in the region. 

Another option is to use an auxiliary Green's function defined by multiplying the original Green's function by~${\rm e}^{-\lambda_1 t}$:
\be
G^{\rm R}_{ij}(t) = {\rm e}^{-\lambda_1 t} G_{ij}(t) \,,
\label{eqn:auxgreen}
\ee
which amounts to a redefinition of the transition matrix:~$M_{\rm R} \equiv M - \lambda_1\mathds{1}$. The eigenvalues are shifted,~$\lambda_\ell \rightarrow \lambda_\ell -\lambda_1$,
but the eigenvectors~$v^{(\ell)}_i$ are unchanged. Similarly we can define an auxiliary first-passage probability density
\be
F_{ij}^{\rm R}(t) = {\rm e}^{-\lambda_1 t}F_{ij}(t) \,,
\ee
with Laplace transform~$F_{ij}^{\rm R}(s) = F_{ij}(s + \lambda_1)$. Since the auxiliary process is stationary, Kemeny's constant is obtained as usual:
\be
{\cal T} = \frac{1}{N}\sum_i v^{(1)\,2}_i \langle t_{ij}^{\rm R} \rangle =\frac{1}{N} \sum_{\ell \geq 2} \frac{1}{|\lambda_\ell - \lambda_1|}\,.
\label{Taux}
\ee

If there is a huge hierarchy between transition rates and leakage rate, such that~$|\lambda_1| \ll |\lambda_{\ell \ge 2} |$, the difference between the two definitions,~\eqref{average MFPT 3} and~\eqref{Taux}, is negligible. It should be emphasized that the modified Kemeny's constant~${\cal T}$ is a meaningful quantity only when~$|\lambda_1| \ll |\lambda_{\ell \ge 2} |$ since the auxiliary Green's function~\eqref{eqn:auxgreen} does not conserve probability as $\sum_{i}\left(M_{\rm R}\right)_{ij} \neq 0$ in general. The largest eigenvalue governs the leakage rate only when the leakage time is much longer than most of the transitions therefore in such situation the subtraction~$|\lambda_\ell - \lambda_1|_{\ell \ge 2}$ has a meaning of mixing rates among the landscape.

\subsection{Dynamical criticality}
\label{optimal}

Adopting~\eqref{average MFPT 3} for Kemeny's constant, this expression reveals a trade-off between minimizing the MFPT and minimizing decay rates. Decay rates cannot be made
arbitrarily small, for otherwise they will result in a large search time. To see this tension in detail, following~\cite{Khoury:2019yoo} we will express~\eqref{average MFPT 3}
as a statistical average over possible realizations of the region. 

In general, the decay rate~$\kappa_i$ of a metastable vacuum depends on its potential energy~$V_i$ as well as the shape of the surrounding potential barriers. The dependence on~$V_i$ comes from various sources. An obvious source is the 4-volume~$H^{-4} \sim V^{-2}$ of the false dS vacuum. Another source is a systematic dependence on
the number of possible destination vacua as a function of~$V$, as expected in a funneled landscape. Lastly, gravitational corrections to the rate per unit volume depend on~$V$, and are particularly significant for vacua near~$M_{\rm Pl}$. In any case, denoting collectively by~$\theta_i$ the parameters characterizing the potential barriers, we can write 
\be
\kappa_i = \kappa(V_i,\theta_i)\,.
\ee
Furthermore, let us denote by~$\rho(V,\theta)$ the underlying joint probability distribution for a given vacuum to have potential energy~$V$ and barrier parameters~$\theta$. In the limit~$N \gg 1$, the mean lifetime of dS vacua can be approximated by a statistical average: 
\be
{\cal T} \simeq \int  {\rm d}V{\rm d}\theta\, \kappa^{-1} (V,\theta) \,\rho(V,\theta)\,.
\ee
Assuming for simplicity that the absolute height of a vacuum and the shape of the surrounding potential barriers are uncorrelated,~$\rho(V,\theta) \equiv \rho(V) \hat{\rho}(\theta)$, this simplifies to
\be
{\cal T} \simeq  \int  {\rm d}V\, \kappa^{-1}(V) \rho(V)\,,
\label{MFPT cont kappa}
\ee
where~$\kappa^{-1}(V) \equiv \int {\rm d}\theta\, \kappa^{-1} (V,\theta)\hat{\rho}(\theta)$ is the average lifetime of vacua with potential energy~$V$. 
Provided~$\rho(V)$ falls off sufficiently fast for large~$V$, the behavior of~${\cal T}$ is controlled by the small-$V$ portion of the integral.  
Making the usual assumption that~$\rho(V)$ is approximately uniform for~$V \ll M_{\rm Pl}^4$~\cite{Weinberg:2000qm},~\eqref{MFPT cont kappa}
simplifies to
\be
{\cal T} \simeq \int_{V_{\rm min}} \frac{{\rm d}V}{M_{\rm Pl}^4 \kappa(V)} \,.
\label{MFPT V}
\ee
The cutoff~$V_{\rm min}$ corresponds to the smallest positive vacuum energy achieved in the region. For a uniform distribution, it is of order
\be
V_{\rm min} \sim \frac{M_{\rm Pl}^4}{N}\,. 
\label{Vmin}
\ee

We are interested in the scaling of~${\cal T}$ with~$N$, in particular whether or not~${\cal T}$ diverges as~$N \rightarrow \infty$. Clearly this is controlled by the behavior of~$\kappa(V)$ as~$V\rightarrow 0$. To maximize the lifetime of dS vacua, as required by the measure,~$\kappa(V)$ should vanish as~$V\rightarrow 0$, but it should do so sufficiently slowly that~${\cal T}$ remains finite or, at worst, diverges slowly as~$N \rightarrow \infty$. To illustrate this point, suppose for concreteness that~$\kappa(V)$ tends to a power law for~$V \ll M_{\rm Pl}^4$ , 
\be
\kappa(V) = \left(\frac{V}{M_{\rm Pl}^4}\right)^{1+ \alpha}\,;\qquad  \alpha \geq -1\,.
\label{kappa dS}
\ee
The lower bound on~$\alpha$ implies that~$\kappa(V) \rightarrow 0$ as~$V \rightarrow 0$, as favored by the measure. (Note that~$\alpha < -1$ is forbidden, as it would entail that low-lying vacua have a lifetime shorter than one e-fold, which is inconsistent.) Substituting this functional form into~\eqref{MFPT V}, the average MFPT is
\be
{\cal T} \sim \left\{\begin{array}{ccl}
\text{constant}  & ~~\text{for} & - 1 \leq \alpha < 0  \\
\log N  & ~~\text{for} &~~~~~\alpha = 0  \\
N^{\alpha} & ~~\text{for} &~~~~~\alpha > 0 \,.
\end{array}\right.
\label{comp scaling}
\ee
Thus for a given~$N \gg 1$ the average MFPT exhibits a sharp growth around the critical value~$\alpha = 0$, which corresponds to
\be
\kappa_{\rm crit} (V) \sim \frac{V}{M_{\rm Pl}^4}\,. 
\label{kappa crit}
\ee
This case delineates between convergent and divergent average MFPT, and as such signals {\it dynamical criticality}. A similar dynamical phase transition
occurs in quenched disordered media, when the probability distribution for waiting times reaches a critical power-law~\cite{disorderedmedia}. 

From a computational perspective, the critical case~\eqref{kappa crit} represents a {\it computational phase transition}~\cite{compPT1,compPT2} in how~${\cal T}$ scales with the effective moduli space dimensionality~$D \sim \ln N$. As mentioned earlier, the average MFPT in generic/glassy regions of the landscape is expected to diverge exponentially
with~$N$, and therefore doubly-exponentially in~$D$, which is consistent with the~\textsf{NP}-hard complexity class of the general problem~\cite{Denef:2006ad}.
The critical case corresponds to~${\cal T}$ scaling as~$\log N \sim D$, and therefore polynomially in~$D$. Thus optimal regions correspond to special, polynomial-time
instances of the general problem. 

It is worth stepping back to highlight the physical significance of~${\cal T}_{\rm crit} \simeq \log N$. As mentioned at the beginning of Sec.~\ref{Kemeny constant},
if our sole concern were to minimize the MFPT to individual low-lying vacua, then the answer would trivially be that the shortest possible MFPT is of order
of a few e-folds, that is,~$\sim {\cal O}(N^0)$. But clearly this would require other vacua in the region to have a short lifetime of a few e-folds,
which is sub-optimal. The ideal compromise is achieved with an average MFPT of~${\cal O}(\ln N)$, at least for low-lying vacua~($V \ll M_{\rm Pl}^4$)
in the region. (Vacua with potential energy close to the fundamental scale can of course have much shorter MFPT.)

Translated to proper time, the critical decay rate~\eqref{kappa crit} corresponds to vacua having an average lifetime given by their dS Page time~\cite{Khoury:2019yoo}:\footnote{With black holes, the Page time is defined as the half-evaporation time~\cite{Page:1993wv}. Its dS analogue is obtained by the Schwarzchild radius with the Hubble radius, and is motivated by the well-known similarities in causal and thermodynamical properties between black hole and dS geometries, {\it e.g.},~\cite{Danielsson:2002td,Danielsson:2003wb,Ferreira:2016hee,Ferreira:2017ogo}.}
\be
\tau_{\rm crit} \sim \frac{M_{\rm Pl}^2}{H^3}\,.
\label{tau crit}
\ee
In the context of slow-roll inflation, it has been shown that the dS Page time coincides with the phase transition to slow-roll eternal inflation~\cite{Creminelli:2008es,Dubovsky:2008rf,Dubovsky:2011uy}. It also implies an upper bound on the maximum number of e-folds that can be described semi-classically~\cite{ArkaniHamed:2007ky}. In pure dS space, it is interesting to note that the coincident graviton 2-point function,~$\langle h^2(x) \rangle \sim H^3 \tau /M_{\rm Pl}^2$, becomes~${\cal O}(1)$ around the Page time, signaling a breakdown of perturbation theory. In any case, if our vacuum is part of an optimal region, then, taking the observed CC as an input, this would predict an optimal lifetime for our vacuum of~$M_{\rm Pl}^2/H_0^3 \sim 10^{130}~{\rm years}$. As mentioned earlier, this agrees to within~$\gsim\; 2\sigma$ with the Higgs metastability result~\cite{Andreassen:2017rzq}:~$\tau_{\rm SM} = 10^{526^{+409}_{-202}}~{\rm years}$.

\subsection{Non-anthropic selection of a small cosmological constant}

In this Section we argue that the early-time measure can favor vacua with the smallest positive CC in the ensemble of regions, thereby possibly offering a non-anthropic solution to the CC problem. 

First, consider the measure~\eqref{measure dS} for dS vacua. As argued in Sec.~\ref{funnel sec}, the ever-hitting probability~${\cal P}_{i}^\infty$ in the downward approximation approaches unity in funneled regions of the landscape, where each dS vacuum has at least one allowed downward transition. Meanwhile, as argued in Sec.~\ref{optimal}, the average MFPT in critical regions of the landscape is~${\cal O}(\ln N)$ for vacua with~$V \ll M_{\rm Pl}^4$. Therefore, within the ensemble of optimal regions, it is clear that the measure is most sensitive to the decay rate~$\kappa_i$. Marginalizing over barrier parameters, as we did in Sec.~\ref{optimal}, we can consider an average measure as a function of potential energy:
\be
f(V) \sim \frac{1}{\kappa(V){\cal T}}    \qquad (V > 0)\,.
\label{f V>0}
\ee
%
% where the average lifetime~$\kappa^{-1}(V)$ was defined below~\eqref{MFPT cont kappa}. 

Next, consider the measure~\eqref{measure AdS} for AdS vacua. Clearly it is maximized for AdS vacua having a large branching ratio ($T_{ai}\; \lsim\; 1$) to dS vacua in optimal regions such that~${\cal P}_{i}^\infty/\langle t_{i}\rangle$ is optimized. Since we expect realistically that a given AdS vacuum is connected to a few ({\it i.e.},~${\cal O}(N^0)$) dS vacua, we obtain
\be
f(V) \sim \frac{1}{{\cal T}} \qquad (V < 0)\,.
\label{f V<0}
\ee

Ignoring any anthropic selection factor, the probability density~${\rm d}P/{\rm dV}$ to measure vacuum energy~$V$ is in general given by
\be
\frac{{\rm d}P}{{\rm d}V} \sim \rho(V) f(V)\,.
\ee
The first factor~$\rho(V)$ is the underlying probability distribution of vacuum energy. The second factor~$f(V)$ is the cosmological measure, given above,
which encodes the dynamics of eternal inflation. Assuming as before that~$\rho(V)$ is approximately uniform for~$|V| \ll M_{\rm Pl}^4$~\cite{Weinberg:2000qm}, the
vacuum energy probability density is controlled by the measure:
\be
\frac{{\rm d}P}{{\rm d}V} = \frac{{\cal N}}{{\cal T}} \left\{\setstretch{1.5}\begin{array}{ccl}
\kappa^{-1}(V)   & ~~\text{for} & V > 0 \,;  \\
1 & ~~\text{for} & V < 0 \,,
\end{array}\right.
\label{dP dV 1}
\ee
where~${\cal N}$ is a normalization constant. Notice that the distribution is generally asymmetric around~$V = 0$, which traces back to the asymmetry of lifetimes: AdS vacua crunch within an e-fold, whereas dS vacua can be long-lived as~$V\rightarrow 0$. Moreover, as mentioned earlier, to maximize the lifetime of low-lying dS vacua~$\kappa(V)$ should vanish as~$V\rightarrow 0$,
hence~${\rm d}P/{\rm d}V$ diverges as~$V\rightarrow 0^+$. Depending on the rate at which~$\kappa(V)$ vanishes, this can favor a small and positive CC, as we now explain.\footnote{Some authors~\cite{Sumitomo:2012wa,Sumitomo:2012vx,Sumitomo:2012cf,Danielsson:2012by,Sumitomo:2013vla,Tye:2016jzi} have suggested that the underlying distribution~$\rho(V)$ may itself peak around~$V = 0$. If so, this would reinforce this feature of~${\rm d}P/{\rm dV}$.}

Consider the power-law dependence~\eqref{kappa dS},~$\kappa(V) = (V/M_{\rm Pl}^4)^{1+\alpha}$, with~$\alpha \geq 0$. 
The probability distribution~\eqref{dP dV 1} is regularized near~$V = 0$ by having a finite number~$N_{\rm tot}$ of vacua in our ensemble. 
To be clear,~$N_{\rm tot}$ is the number of vacua in {\it all} optimal funneled regions combined. Akin to~\eqref{Vmin}, this sets a low potential energy cutoff: 
\be
\Lambda_{\rm min} \sim \frac{M_{\rm Pl}^4}{N_{\rm tot}}\,.
\ee
We also impose a high energy cutoff of~$\Lambda_{\rm max}  \sim M_{\rm Pl}^4$, at which scale the assumption of a uniform distribution and/or the semi-classical approximation breaks down. The normalized distribution for~$\alpha > 0$ is then 
\be
\frac{{\rm d}P}{{\rm d}V} = \frac{1}{M_{\rm Pl}^4{\cal T}_\alpha}
\left\{\setstretch{1.5}\begin{array}{ccl}
\left(\frac{M_{\rm Pl}^4}{V}\right)^{1+\alpha}  & ~~\text{for} & \Lambda_{\rm min} \leq V \leq M_{\rm Pl}^4   \\
1 & ~~\text{for} & -M_{\rm Pl}^4 \leq V \leq -\Lambda_{\rm min} 
\end{array}\right. \,,
\label{dP dV alpha < 0}
\ee
where~${\cal T}_\alpha \simeq \alpha^{-1} \left(\frac{M_{\rm Pl}^4}{\Lambda_{\rm min}}\right)^\alpha$ is the average MFPT. It is easy to see that the probability of measuring a {\it negative} vacuum energy is utterly negligible. Instead, the~predicted~95\% confidence range lies almost entirely on the positive side:
\be 
1 \;\lsim\; \frac{V}{\Lambda_{\rm min}} \;\lsim\; 20^{1/\alpha}\,.
\ee
This can account for the observed CC of~$\simeq 10^{-122}\,M_{\rm Pl}^4$ if the total number of vacua in the ensemble
of funneled regions is within the range~$1 \;\lsim\; 10^{-122}N_{\rm tot} \;\lsim\; 20^{1/\alpha}$.  Note that the proposed solution to the CC problem breaks down in the limit~$\alpha \rightarrow 0$. 
The probability distribution for the critical case~$\alpha = 0$ still favors a positive CC, but it is only log-uniform. 

It seems that we are required to have a huge number of vacua , $N_{\rm tot}> 10^{122}$. We stress that our argument is very different from two well known stories. The first one assumes a uniform distribution of cosmological constant and, as long as there is a large enough number of vacua, our vacuum is anthropic. We are certainly not involving any anthropic principle in our analysis. The second well known story is mediocrity in which the distribution of vacuum energy, without considering any dynamics, peaks at zero. The distribution derived here folded in early dynamics which takes into account accessibility of landscape regions. It means that vacua of small CC vacua are easily accessed early on but not necessarily dominate the landscape.

\section{Early-time measure in Bousso-Polchinski flux landscapes}
\label{BP section}

It is known that the string theory landscape is not entirely random. There are well-known regular structures, such as vacua generated by flux compactification \cite{Giddings:2001yu}. We will use the Bousso-Polchinski (BP) landscape~\cite{Bousso:2000xa} as an illustration to capture some of the essence of the discussion in previous Sections.

The BP landscape is a field theory landscape with many vacua coming from flux compactifications. The effective potential energy of each vacuum is given by 
\begin{equation}
 V_{\rm eff} = \Lambda_0 + \frac{1}{2} \sum_{i=1}^D g_i^2 n_i^2\,;   \quad n_i \in\mathds{Z}\,,
\end{equation} 
where~$g_i$ is the charge of the~$i$-th quantum,~$n_i$ is the number of flux units, and~$D$ is the total number of flux directions. Therefore, the lowest-lying set of vacua form a funnel. 
The BP flux landscape is a valid effective description below a certain scale, which translates to a field range within a radius
\be
|\phi|\leq R_{\rm max} \,.
\ee
This field range roughly corresponds to~$V_{\rm eff}$ remaining below the string scale. Beyond this field range one expects that backreaction on the internal geometry and other effects may take over. 

A flux transition between vacua, such that~$n_i \to n_i +\Delta n_i$, is caused by brane nucleation. The nucleation rate is~$\sim {\rm e}^{-B}$. In the thin-wall, probe-brane approximation and ignoring gravitational effects,~$B$ is given by 
\be
B \sim \frac{\sigma^4}{\left(\Delta V\right)^3} = \frac{8\sigma^4}{g^6 \Big[\sum_{i=1}^D\Big(n_i^2 - (n_i + \Delta n_i)^2\Big)\Big]^3} \,,
\ee
where the brane tension is approximated by
\begin{equation}
 \sigma \approx M_4 \left( \sum_{i=1}^D g_i^2 \Delta n_i^2\right)^{1/2}\,.
\label{sigma BP}
\end{equation}
The scale~$M_4$, of order~$M_{\rm Pl}$, will be chosen such that the exponent~$B$ can be handled numerically. In an effective potential describing the BP landscape, vacua are located at the lattice sites~$\vec{\phi} = \{ g_1 n_1 , \dots, g_Dn_D  \}$. Thus we see that the brane tension is proportional to the field-space distance distance between vacua,~$\big\vert\Delta \vec{\phi}\big\vert = \left( \sum_i^D g_i^2 \Delta n_i^2\right)^{1/2}$. 

In the numerical experiments below, we study a BP landscape region in~$D =1,2,3$, with  
\be
\Lambda_0 =  10^{-2} M_{\rm Pl}^4\,;\qquad g_i^2 = g^2 = 2\times 10^{-6} M_{\rm Pl}^4\,. 
\ee
Notice that all charges~$g_i$ are the same, for simplicity, since qualitatively the lattice spacing along different directions does not affect the quantities of interest. 
The region is taken to be a ball of radius~$R < R_{\rm max}$, with 
\be
R = 8 g\,.
\ee
The environment surrounding the region consists of all flux vacua in the shell~$R \leq \big\vert\vec{\phi}\big\vert \leq R_{\rm max}$. To account for leakage into the environment,
we include in the total decay rate of each vacuum in the region a contribution from flux transitions to exterior vacua in the shell~$8g \leq  \big\vert\vec{\phi}\big\vert \leq 10 g$. (We have checked that  
further enlarging the shell has negligible effect on the rates within the region.) Lastly, we choose the scale~$M_4$ in~\eqref{sigma BP} such that the wall tension for unit-flux transition is~$\sigma = 10^{-4} M_{\rm Pl}^3$.

\subsection{Mean residency time and Kemeny's constant}

A first quantity of interest is the mean residency time~$\langle t_{\rm esc}\rangle$, or mean escape time, of the funnel region, defined in~\eqref{Tesc}.
Recall that our measure prefers a stable funnel with long residency time. As illustrated in the left panel of Fig.~\ref{Tesc fig},~$\langle t_{\rm esc}\rangle$ {\it decreases with
field-space dimension~$D$} for regular lattices like the BP flux vacua.

\begin{figure}[h]  
\center
 \includegraphics[scale=0.28]{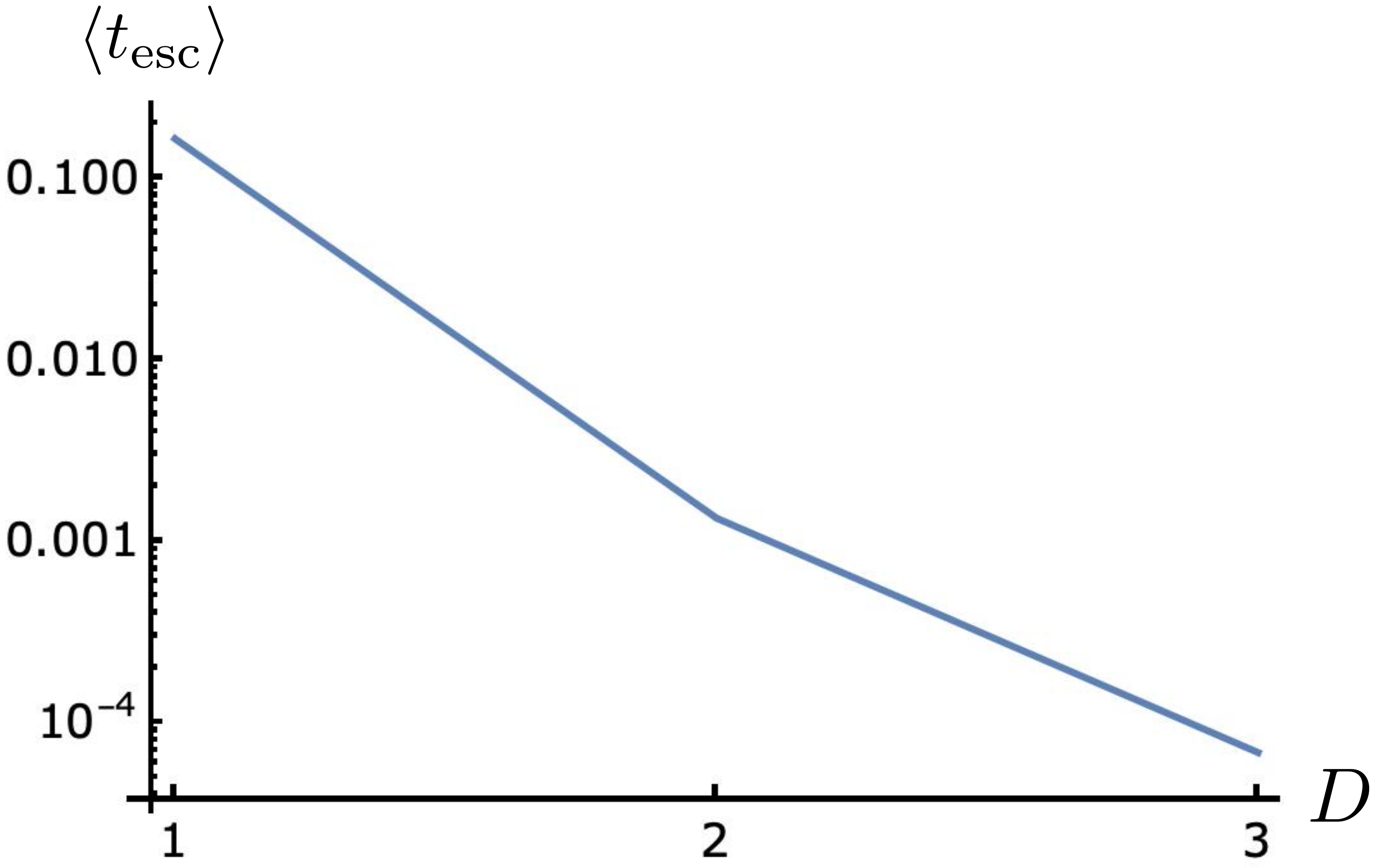}
 \includegraphics[scale=0.31]{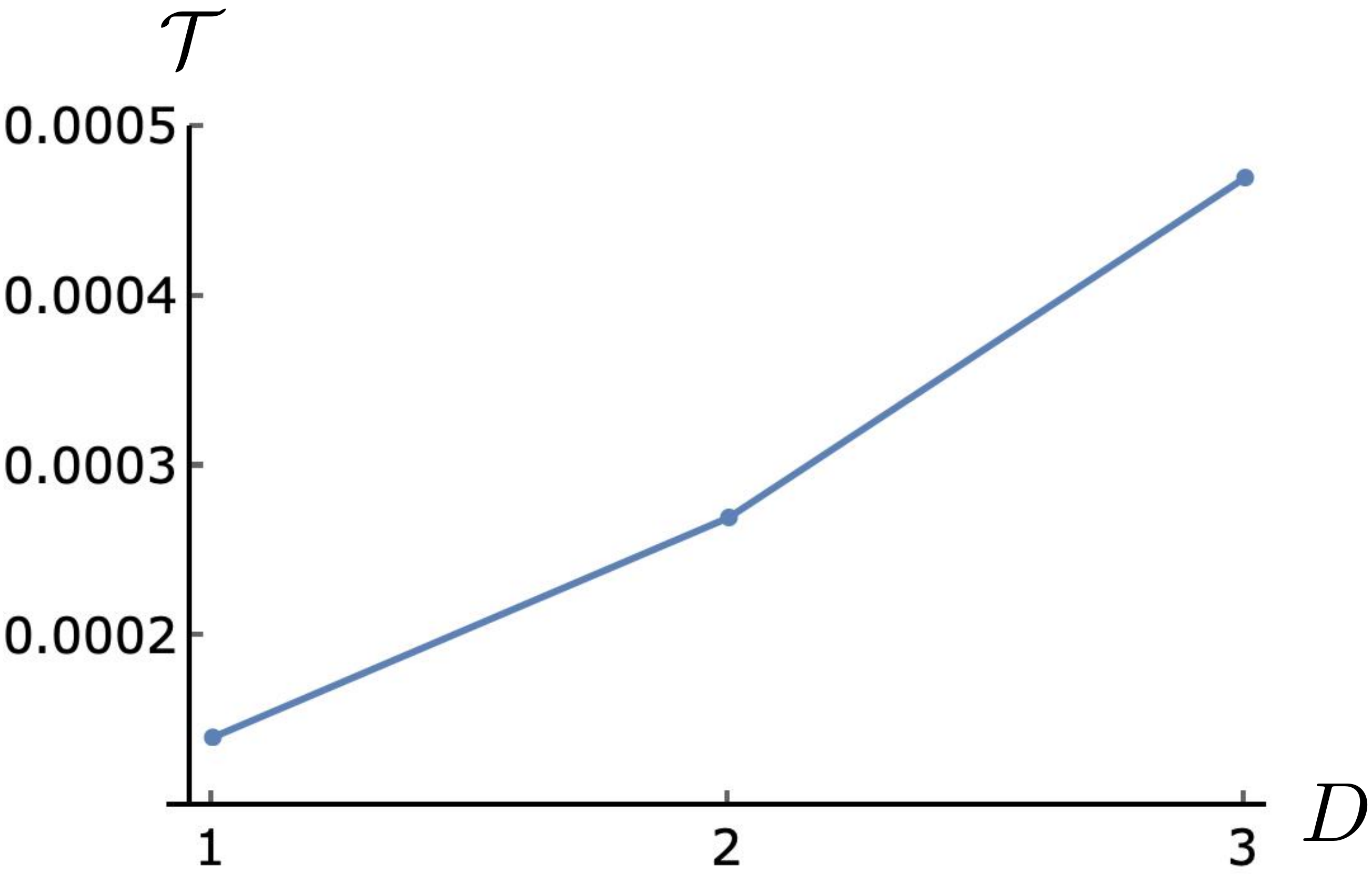}
\caption{The mean residency time~$\langle t_{\rm esc}\rangle$ (left panel) and Kemeny's constant~${\cal T}$ (right panel) as a function of the number~$D$ of fluxes in a BP landscape region, described in the main text. The decrease in~$\langle t_{\rm esc}\rangle$ and the growth in~${\cal T}$ with increasing number of fluxes both favor small~$D$.}
\label{Tesc fig}
\end{figure}

There are two reasons behind the drop in escape time. The first reason is that adding an extra flux direction introduces new up-tunneling paths of relatively less suppressed rate. To see this, recall from~\eqref{detailed balance} that the up-tunneling rate from a true (t) vacuum to a false vacuum (f) is related to downward rate by 
\begin{equation}
\frac{\kappa_{{\rm t}\to {\rm f}}}{\kappa_{{\rm f}\to {\rm t}}}\sim \exp\left[ 24 \pi^2 M_{\rm Pl}^4 \left( \frac{1}{V_{\rm t}} - \frac{1}{V_{\rm f}} \right) \right]\,.
\label{detailed balance BP}
\end{equation}
(Since~$V \ll M_{\rm Pl}^4$ within the funnel region, the exponential factor is the dominant suppression factor for up-tunneling.) Consider, for instance, up-tunneling from~$(n_1, n_2) = (4,0)$ to~$(5,0)$. In one dimension where~$n_2$ is frozen, the exponential factor in~\eqref{detailed balance BP} is huge. When transitions in~$n_2$ are allowed, however, the zigzag-like path~$(4,0) \to (4,1) \to (3,3) \to (4,2) \to (5,0)$ is significantly less suppressed, since the exponential factor is much smaller as the climb in potential is reduced in each step.  

The second reason for the decline in escape time is due to the fact that there are many more of these zigzag paths when extra flux directions are introduced. The phenomenon we observe here actually matches that of a diffusion problem. It is well known that a random walker has higher tendency to diffuse outward from the origin in high dimensions.\footnote{A classic result in random walk theory is P\'olya's theorem~\cite{polya}, which states that simple random walks on the integer lattice~$\mathds{Z}^d$ are recurrent (the ever-return probability to the starting node is~1) for~$d \leq 2$, and transient for~$d > 2$ (the ever-return probability is~$< 1$).}

The right panel of Fig.~\ref{Tesc fig} shows the modified Kemeny's constant, defined in~\eqref{Taux}. (Because there is a large hierarchy between transition rates and leakage rate, {\it i.e.},~$|\lambda_1| \ll |\lambda_{\ell \ge 2} |$, the difference between~\eqref{average MFPT 3} and~\eqref{Taux} is negligible.) We see that Kemeny's constant, which quantifies the mixing time within the region, increases {\it slowly} with~$D$. Importantly, it does not grow exponentially, which reflects the ordered/funnel nature of the BP region.

\subsection{Ever-hitting probability, MFPT, and the new measure}

Next we consider the new measure~\eqref{measure dS}. As initial conditions, we choose for concreteness the initial distribution to be given by the dominant eigenvector of~$M$ associated with the largest eigenvalue~$\lambda_1$:
\be
p_i = v_{M\;i}^{(1)}\,.
\label{pi dom}
\ee
Recall that~$v_M^{(1)}$ peaks for the longest-lived dS vacuum in the region, which in the BP lattice coincides with the origin~$\vec{n} = 0$. Thus, with these initial conditions, accessing vacua further up in the funnel requires up-tunneling. Below we will check that our results are fairly insensitive to the choice of initial distribution.

Figure~\ref{fig:MPFTBP} shows various observables for the BP lattice, plotted along a central line~$(n_1,0,0)$ for different number of fluxes~$D$. We see that the
ever-hitting probability~${\cal P}^{\infty}_i$ (top left panel) and weighted MFPT~$\langle t_i \rangle$ (top right panel) both decreas with increasing~$D$. Both trends are consistent with the aforementioned growth in the number of up-tunneling paths with smaller suppression with increasing~$D$. In higher~$D$, random walks can climb the funnel more quickly, resulting in shorter MFPT\footnote{Note that this is not in contradiction with the slow growth in Kemeny's constant shown in~Fig.~\ref{Tesc fig}. The point is that~$\langle t_i \rangle$ is defined in~\eqref{average t_i} with an additional weighing factor of~${\cal P}^{\infty}_i$. The decrease in~${\cal P}^{\infty}_i$ with~$D$ overwhelms the slow growth in the pairwise MFPT's~$\langle t_{ij} \rangle$.}, but there is a greater probability of escape, resulting in a reduced ever-hitting probability. Notice that, whereas the MFPT to vacua near the boundary decreases sharply with~$D$, the MFPT to the vicinity of the bottom only slightly changes with~$D$, as expected given our initial conditions. 

\begin{figure}[h]  
\center
 \includegraphics[scale=0.29]{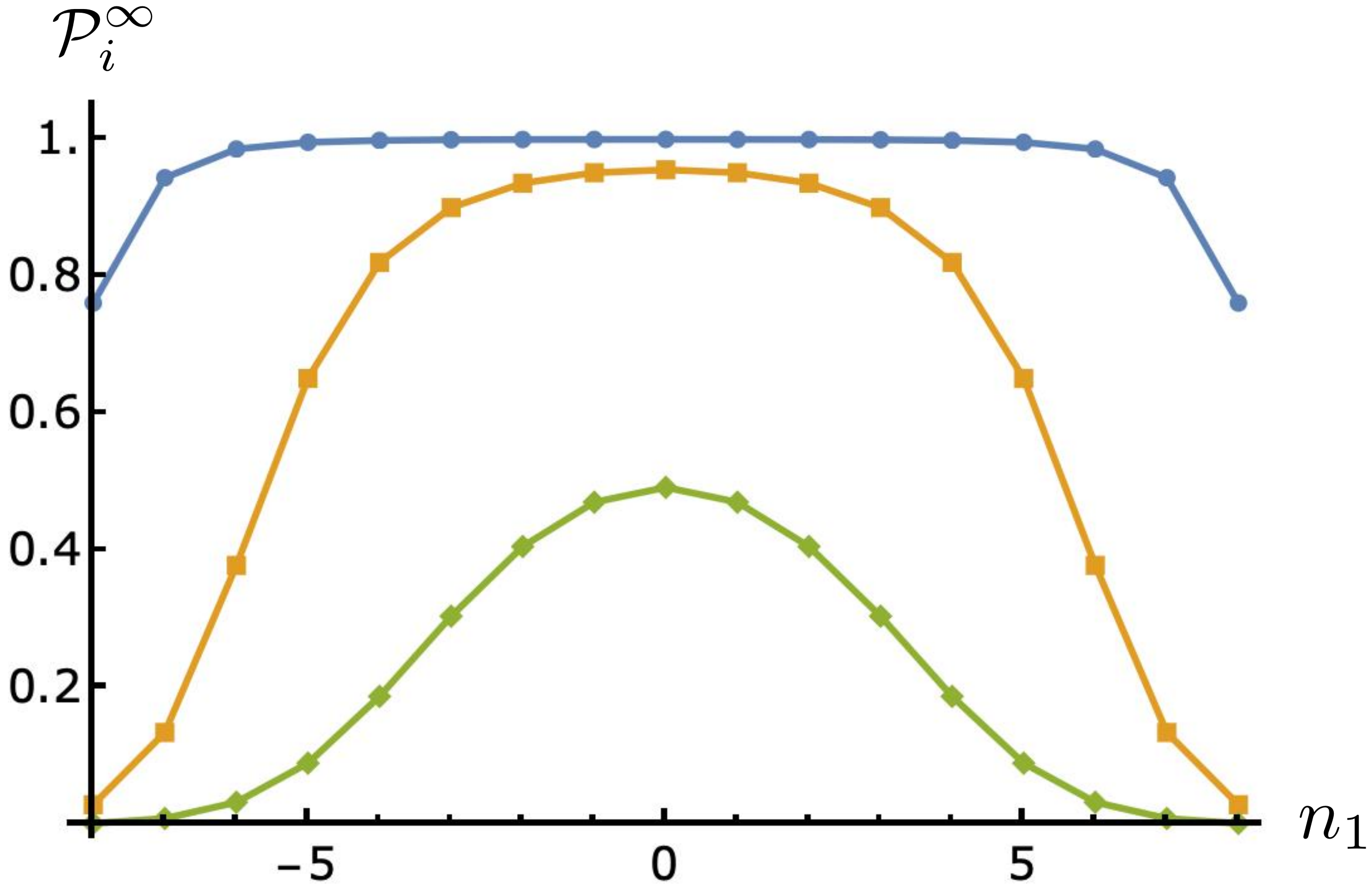}
 \includegraphics[scale=0.29]{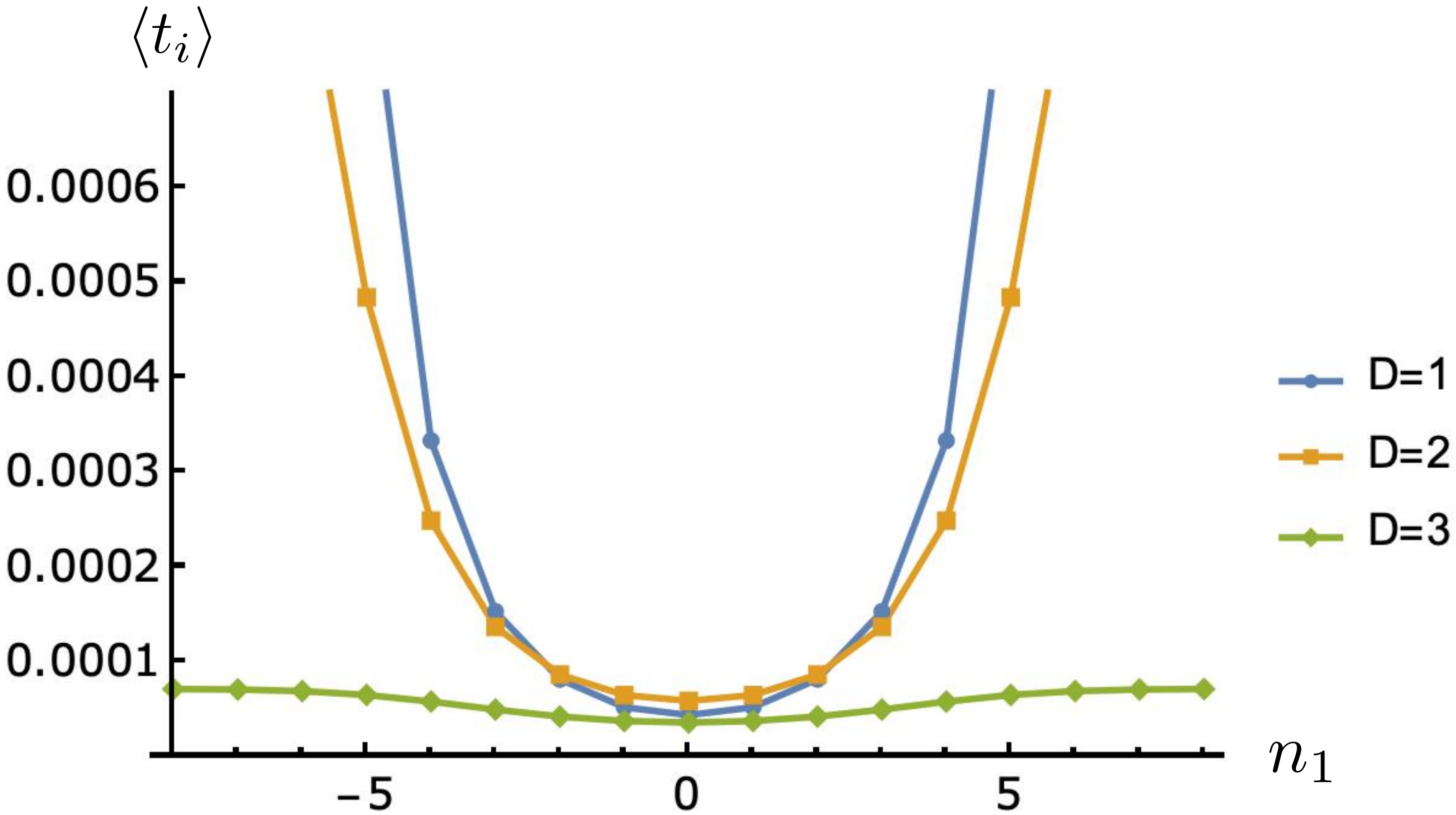}\\
\vspace{0.3cm}
   \includegraphics[scale=0.29]{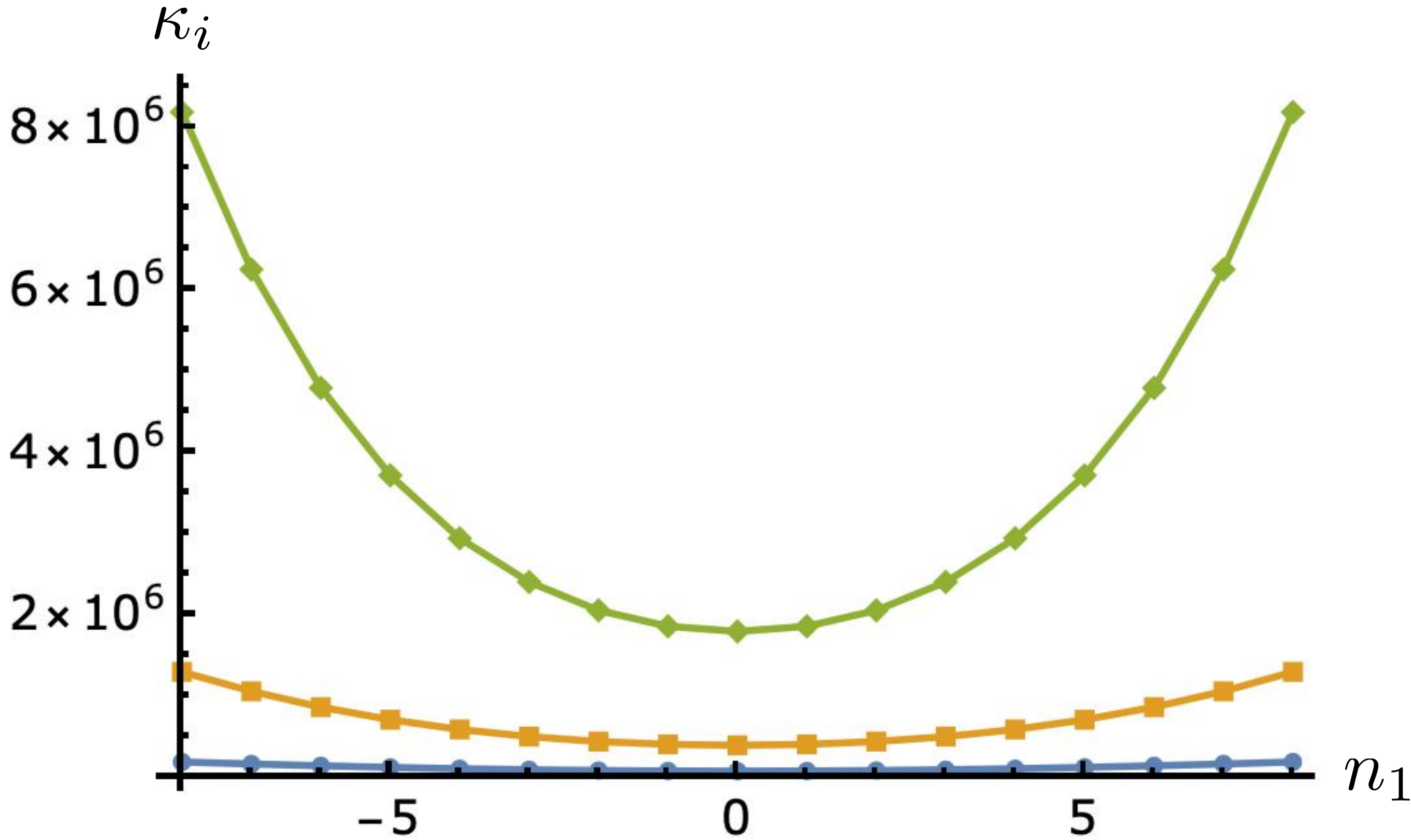}
 \includegraphics[scale=0.29]{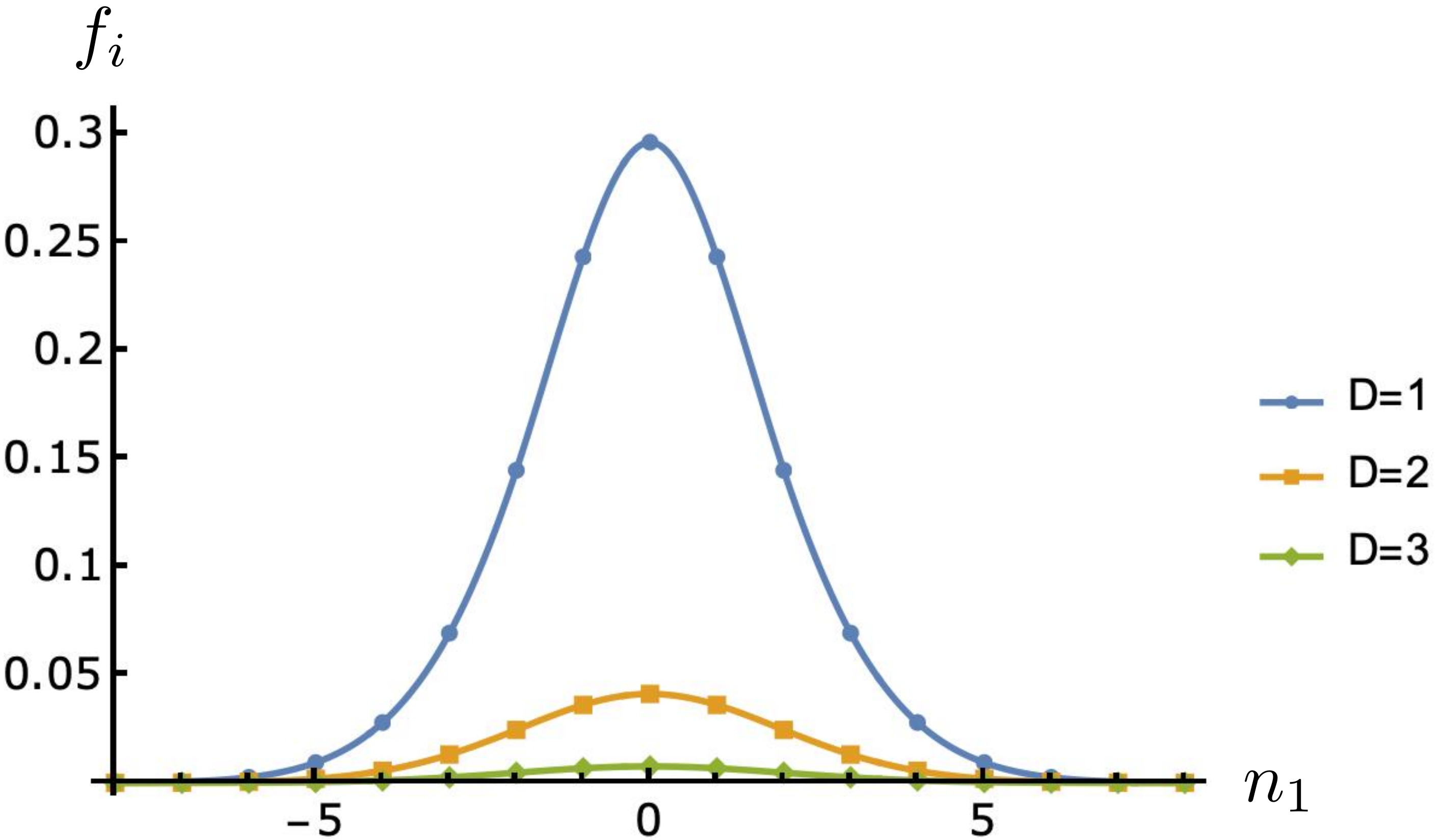}
\caption{The weighted ever-hitting probability~${\cal P}^{\infty}_i$ ({\it top left}), weighted MFPT~$\langle t_i \rangle$ ({\it top right}), decay rate~$\kappa_i$ ({\it bottom left}), and early-time measure~$f_i$ ({\it bottom right}) for a few values of~$D$, both plotted along the line~$(n_1,0,0)$.}
\label{fig:MPFTBP}
\end{figure}

Meanwhile, the decay rates~$\kappa_i$ of different vacua (bottom left) grow with increasing~$D$, as expected given the increasing number of decay channels in higher~$D$. These factors combine to give an early-time measure~$f_i$ (bottom right), defined in~\eqref{measure dS}, which clearly {\it decreases with increasing~$D$.} Early dynamics on BP landscapes therefore prefer regions with low effective dimensionality.   

We have checked that these conclusions do not strongly depend on initial conditions. Figure~\ref{BP compare fig} compares the ever-hitting probability (left) and MFPT (right) in~$D = 2$ for the original, ``dominant eigenvector" distribution (blue curves), which peaks in the center, and an alternative distribution~$p_i \sim {\rm e}^{\vec{n}^2/16}$ (orange curves), which peaks on the boundary. Both ever-hitting probability and MFPT only show a small variation. This confirms that our results are insensitive to the choice of initial distribution.

\begin{figure}[h]  
\center
 \includegraphics[scale=0.29]{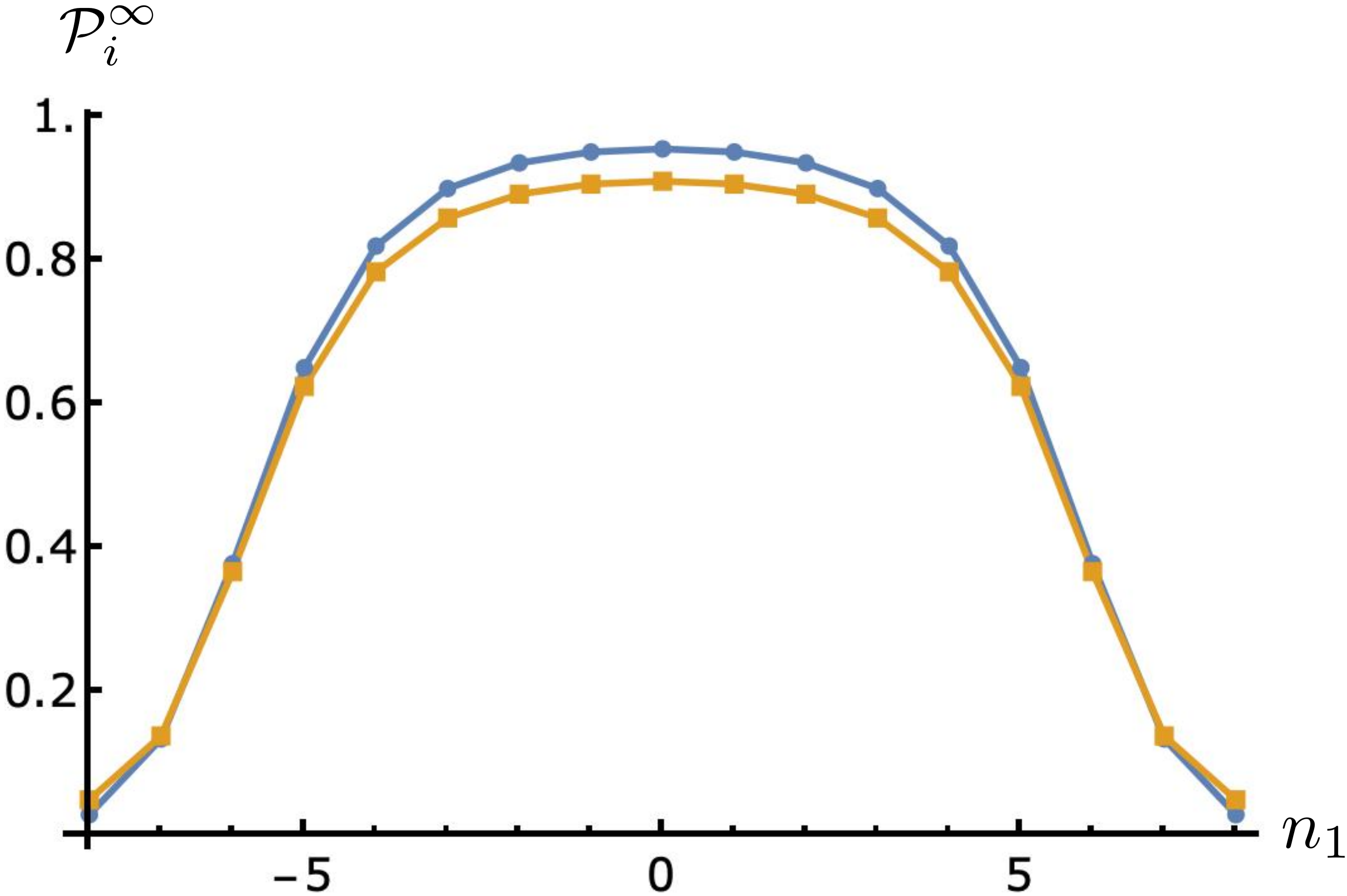}
 \includegraphics[scale=0.31]{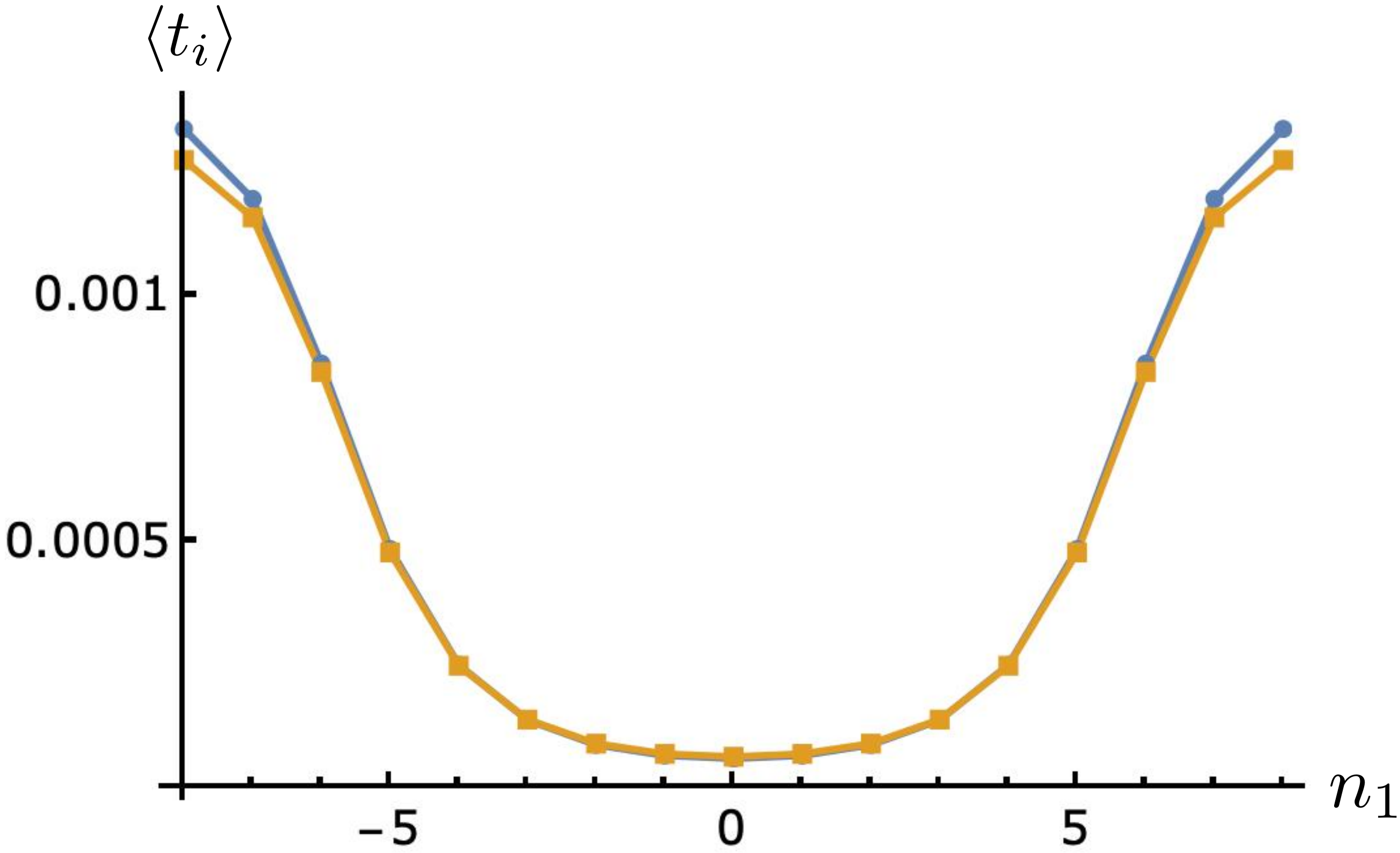}
\caption{The ever-hitting probability~${\cal P}^{\infty}_i$ ({\it left}) and MFPT~$\langle t_i \rangle$ ({\it right}) for the dominant-eigenvector initial distribution~$p_i = v_{M\;i}^{(1)}$ (blue curves), which peaks at the center, and for an alternative distribution~$p_i \sim {\rm e}^{\vec{n}^2/16}$ (orange curves), which instead peaks on the boundary. These are for $D = 2$ and once again plotted along~$(n_1,0,0)$.}
\label{BP compare fig}
\end{figure}

\subsection{Implication of low dimensionality on particle spectrum}

The above analysis has established that regular landscape lattices with fewer field-space dimensions have longer escape time (stability), smaller Kemeny's constant (shorter mixing time), higher ever-hitting probability (accessibility), and overall higher early-time measure. Thus, regions with small field-space dimensionality are accessed quickly, and the space-time volume they occupy grows at a faster rate.  
In what follows we will explore the consequences for the number of light moduli and the effective dimensionality of moduli space.

It is well-known that the string theory landscape is fundamentally of high dimensionality, due to a vast number of moduli fields, such as the complex structure moduli and K\"ahler moduli. It stands to reason, however, that if potential barriers are high enough to suppress tunneling along certain directions, then this would have the effect of reducing the effective field space dimensionality in the low energy theory. Thus it is natural to expect that a large enough tunneling action also imposes constraints on the local curvature of the potential around a vacuum. We demonstrate this below with two familiar potentials, both within a natural range of parameters. 

In order to suppress tunneling to along some field directions, the bounce action should be bounded below,
\begin{equation}
B > B_0\,,
\end{equation}
where~$B_0 \gg 1$. In a high-dimensional moduli space, the bounce solution to a nearby vacuum in general traces a complicated path in field space. Whatever this path may be, it minimizes the  Euclidean action, hence any one-dimensional path (say along a mass eigenstate) will necessarily have higher action, call it~$B_{1{\rm d}}$. That is,
\be
B_{1{\rm d}} > B > B_0\,.
\label{B1 bound}
\ee
The key point is that~$B_{1{\rm d}}$ satisfies the same lower bound as the actual bounce action. Explicitly, in the thin-wall approximation and ignoring gravitational effects,
we have the usual expression,
\be
B_{1{\rm d}} =  \frac{27 \pi^2\sigma_0^4}{2\epsilon^3}\,, 
\ee
where~$\epsilon$ is the difference in vacuum energy, and~$\sigma_0$ is the domain wall tension
\be
\sigma_0 = \left\vert \int_{\phi_{\rm f}}^{\phi_\star} {\rm d}\phi \sqrt{2\left(V(\phi) - V_{\rm f}\right)}\right\vert\,.
\label{sigma0}
\ee
Here~$\phi_\star$ denotes the field value on the other side of the barrier where~$V(\phi_\star) = V_{\rm f}$. Since we generally expect~$\epsilon$ to be of the same order in all field directions, we focus our attention on~$\sigma_0$. 

The first type of one-dimensional potential to consider naturally is the familiar symmetry-breaking potential
\begin{equation}
 V(\phi) = \frac{\lambda}{4}\left( \phi^2 -  \frac{m^2}{2\lambda} \right)^2 + {\cal O}(\epsilon)\,,
\end{equation} 
where~$m$ is the mass at the minimum. The domain wall tension~\eqref{sigma0} in this case gives~$\sigma_0  = \frac{m^3}{3\lambda}$.
Since the coupling~$\lambda$ is typically~$\lsim\; {\cal O}(1)$ number, the lower bound~$B_{1{\rm d}}>B_0$ translates to a lower bound on the
mass along this field direction:
\begin{equation}
 m > \left(\frac{3 \lambda B_0}{\pi^2} \right)^{1/12} \epsilon^{1/4}\,.
 \label{mbound1}
\end{equation}

The second type of single-field potential is the axionic-type potential:
\begin{equation}
 V(\phi) = m^2 f^2 \left(1-\cos \frac{\phi}{f} \right) +{\cal O}(\epsilon)\,.
\end{equation}
where~$m$ and~$f$ are the axion mass and decay constant, respectively. Such potentials naturally appear in low energy theories from string compactifications~\cite{Svrcek:2006yi}, as well as in particle physics to describe pseudo Nambu-Goldstone bosons. The wall tension in this case is~$\sigma_0 = 8 f^2 m$. The lower bound~$B_{1{\rm d}}>B_0$ then implies
\begin{equation}
 m > \frac{1}{8f^2} \left( \frac{\epsilon B_0}{27\pi^2}\right)^{1/4} \;\gsim\; \frac{1}{8M_{\rm Pl}^2} \left( \frac{\epsilon B_0}{27\pi^2}\right)^{1/4}\,,
 \label{mbound2}
\end{equation}
where in the last step we have used the expectation that~$f \;\lsim\; M_{\rm Pl}$ in string theory~\cite{Banks:2003sx}.  

The message is clear. Suppressed tunneling rates, or equivalently large bounce actions, along certain field directions translate to large scalar masses, as~\eqref{mbound1} and~\eqref{mbound2} show. This in turn reduces the effective dimensionality of moduli space. Early dynamics in the landscape favors regions with low effective field space dimensions. As illustrated by the examples above, these regions are likely to have a hierarchical mass spectrum with few light scalars.

\section{Conclusions}

In a situation like eternal inflation where our data is replicated at (infinitely-many) other space-time events, it is necessary to make a prior assumption about our location within the multiverse to extract physical predictions~\cite{Hartle:2007zv,Srednicki:2009vb}. The principle of mediocrity, traditionally assumed in approaches to the measure problem, corresponds to a uniform prior. 
This assumption entails that we live at asymptotically late times in the unfolding of eternal inflation, when the relative occupational probabilities of different vacua has settled to a near-equilibrium distribution. But it is also possible that we are typical of a more restricted class of observers in the multiverse. 

In this paper, building on earlier work	~\cite{Khoury:2019yoo,Khoury:2019ajl,Kartvelishvili:2020thd}, we have studied the implications of a simple, yet powerful, alternative to the principle of mediocrity, namely that we exist during the approach to equilibrium, {\it i.e.}, at times much earlier than the exponentially-long mixing time for the landscape. This is motivated by the near-criticality of our universe, embodied by Higgs metastability and the weak hierarchy and CC problems. Indeed, in the natural world, non-equilibrium (open, dissipative and slowly-driven) systems are generically found at criticality, a phenomenon known as ``generic scale invariance"~\cite{grinstein}. It is natural to envision that the near-criticality of our universe is similarly related to non-equilibrium critical phenomena in landscape dynamics.

In the approach to equilibrium, we are most likely to reside in habitable vacua that are easily accessed under time evolution. This translates to a search optimization problem: accessible vacua reside in optimal regions of the landscape where vacuum transitions are efficient. Using first-passage statistics, which are tailor-made for the problem of interest, we argued that vacua that maximize their volume have maximal ever-hitting probability, minimal MFPT, and minimal decay rate. These three requirements can be succinctly captured by a dimensionless quantity, which we take as a working definition of an early-time measure. 

The idea that we live at early times is a predictive guiding principle, with many phenomenological implications that are distinct from the standard quasi-equilibrium approach. The first implication is that our vacuum lies deep in a funneled region, akin to the folding energy landscapes of proteins. A second implication is that optimal regions are characterized by vacua that are relatively short-lived, with lifetimes of order their dS Page time. For our vacuum, taking the inferred CC as input, this time scale is~$\sim 10^{130}$~years, which is surprisingly consistent with the SM lifetime estimate. This offers a natural explanation for the inferred metastability of the electroweak vacuum. A third implication is the distribution of vacuum energies within funneled regions, which favors small-CC vacua. This can offer a non-anthropic solution to the CC problem, provided there are sufficiently many vacua in the entire ensemble of funnels.

Lastly, by considering the concrete example of BP flux lattices, we uncovered yet another key implication: the early-time measure favors regular landscape lattices with the fewest number of flux dimensions.
In terms of the moduli space, this means that potential barriers must be high enough to suppress tunneling along certain directions, which effectively reduces the field space dimensionality in the low energy theory. 
Thus, this favors compactifications with a large hierarchy between the lightest modulus, controlling transitions between vacua, and all other K\"ahler and complex structure moduli.
 
The framework developed here offers many future directions that we plan to pursue:

\begin{itemize}

\item In a forthcoming paper, we will study the possibility that the approach to equilibrium in landscape dynamics exhibits a directed percolation phase transition~\cite{Odor:2002hk} --- the paradigmatic non-equilibrium critical phenomenon. Preliminary results show that the early-time measure peaks near the percolation threshold, which would imply that optimal regions are poised at directed percolation criticality.

\item Another enticing direction pertains to the effective dimensionality~$d$ of space-time. A straightforward calculation shows that, for fixed potential and Planck scale, the tunneling action for CDL and Hawking-Moss instantons grows with~$d$. Thus, fast transition rates, as favored by the early-time measure, favors small~$d$. This may offer a novel mechanism to explain the low dimensionality of our universe.

\item It will be fascinating to further explore the implications of our residing in a funnel, capitalizing on the vast body of work in protein folding. Indeed, the problem of optimizing the accessibility of low-energy states in complex energy landscapes is a problem that Nature has already solved for us! From a network perspective, it has been shown~\cite{Doye_2002,Rao_Caflisch_2004} that the funneled energy landscapes of simple proteins exhibit properties shared by many other ``real-world" networks, in particular scale-free degree and clustering coefficient distributions, as well as the small-world property. 

\end{itemize}

\vspace{.4cm}
\noindent
%% Anybody else to thank?
{\bf Acknowledgements:} We thank Alan Guth, Guram Kartvelishvili, James Halverson, Eleni Katifori, Cody Long, Minsu Park, Anushrut Sharma, Alex Vilenkin and Nathaniel Watkins for helpful discussions. This work is supported by the US Department of Energy (HEP) Award DE-SC0013528, NASA ATP grant 80NSSC18K0694, and by the Simons Foundation Origins of the Universe Initiative.

\appendix

\section{First-passage statistics}
\label{first-pass app}

In this Appendix we collect some results of first-passage statistics.

\subsection{Single-target first-passage time (MFPT)}

% Assume no leakage and no terminals
We begin with a derivation of an alternative expression for the MFPT,~$\langle t_{ji}\rangle \equiv T_{ji}$, which is valid whenever the region can be approximated as having no leakage and no terminals. That is, this applies for closed, dS-only regions. 

Let~$P^{\rm F}_{ji}(t)$ be the probability that a random walker starts from node~$i$ and visit node~$j$ for the first time at time~$t$. 
Since there is no leakage or terminals, probability is conserved, and it is easy to see that~$P^{\rm F}_{ji}(t)$ can be calculated as 
\begin{equation}
   P^{\rm F}_{ji}(t)= 1- \left[ \hat{e}_{N-1} \,{\rm e}^{M^{(j)} t} \right]_i \,,
\end{equation}
where~$\hat{e}_{N-1}=(1,\dots, 1)$ is the~$(N-1)$ row vector with all elements being~1, and~$M^{(j)}$ is defined as the submatrix of~$M$ without the~$j$-th row and column.
In particular,~$M^{(j)}$ governs all the transitions that do not enter node $j$. Obviously the first passage density is evaluated as $F_{ji}(t) =  \frac{{\rm d}}{{\rm d}t} P^{\rm F}_{ji}(t) $ therefore the MFPT~$T_{ji}$ can be derived as 
\begin{align} 
   T_{ji} = \int_0^{\infty} \rd t   \, t \frac{{\rm d}}{{\rm d}t} P^{\rm F}_{ji}(t)  = -\left[  \hat{e}_{N-1} \left(M^{(j)}\right)^{-1} \right]_i \, .  
\label{eqn:MFPTinversesum}
\end{align}
To be clear, $\hat{e}_{N-1} \left(M^{(j)}\right)^{-1}$ is itself an~$(N-1)$ dimensional row vector, whose~$i^{\rm th}$ component is extracted to obtain~$T_{ji}$. 

\subsection{Two-target first-passage time}

Next, we generalize the above derivation to obtain the two-target search time, once again neglecting leakage and terminals. Let~$P^{\rm F}_{jk,i}(t)$ be the probability that a random walker starting from~$i$ has visited both~$j$ and~$k$ for the first time at time~$t$, with $i\neq j,k$. To find the mean first visit time~$T_{jk,i}$ from~$i$ to the~$(j,k)$ pair, we first consider the probability that the random walker has never visited~$(j,k)$ at time~$t$. It is given by 
\begin{equation}
  \sum_{l}\left[ e^{M^{(jk)} t}\right]_{li} =\left[ \hat{e}_{N-2}\, {\rm e}^{M^{(jk)} t}\right]_i \,,
\end{equation}
where~$\hat{e}_{N-2}=(1,\dots, 1)$ is the~$(N-2)$ row vector with all unit entries, and~$M^{(jk)}$ is the submatrix of~$M$ without the~$j$-th and~$k$-th rows and columns. Therefore, from the vein diagram~$P^{\rm F}_{jk,i}(t)$ is given by 
\be
P^{\rm F}_{jk,i}(t) = P^{\rm F}_{ji}(t) +P^{\rm F}_{ki}(t) + \left[ \hat{e}_{N-1}\, {\rm e}^{M^{(jk)} t}\right]_i -1.
\ee
Then~$T_{jk,i}$ is easily evaluated,
\bea
\nonumber
 T_{jk,i} &=& - \left[ \hat{e}_{N-1} \left(M^{(j)} \right)^{-1} \right]_i - \left[ \hat{e}_{N-1} \left(M^{(k)}\right)^{-1}  \right]_i + \left[ \hat{e}_{N-2} \left(M^{(jk)}\right) ^{-1} \right]_i  \\
  &=&  T_{ji} + T_{ki}+ \left[ \hat{e}_{N-2} \left(M^{(jk)}\right)^{-1}  \right]_i\,.
\label{Tjki}
\eea
One should keep in mind that~$i\neq j, k$.
 
Without loss of generality, let us set~$j=1$  and~$k=2$. It turns out~$T_{12,i}$ can be written solely in terms of the single-target MFPTs~$T_{1i}$,~$T_{2i}$,~$T_{12}$ and~$T_{21}$:
\begin{align}
\boxed{T_{12,i} =  \frac{T_{1i}T_{12} + T_{2i}T_{21} + T_{12}T_{21} }{T_{12}+ T_{21}}\,.}
\label{eqn:Trecur}
\end{align}
It is identical to the result of discrete time Markov process \cite{doi:10.1063/1.4990866} but the proof from discrete time version does not apply directly to the continuous time Markov process. To prove this, first rearrange the identity in the following form
\begin{equation}
 T_{12,i}\big( T_{12}+ T_{21}\big)=T_{1i}T_{12} + T_{2i}T_{21} + T_{12}T_{21} \,,
\end{equation}
which means, using~\eqref{Tjki}, that what we need to show is 
\be
T_{1i}T_{21} + T_{2i}T_{12}  + \big(T_{12}+ T_{21}\big) \left[ \hat{e}_{N-2}  \left(M^{(12)}\right) ^{-1}  \right]_i = T_{21}T_{12}\,.
\label{eqn:Trecur equiv}
\ee

To do so, write~$M^{(1)}$ and~$M^{(2)}$ explicitly as
\be
M^{(1)} = \begin{pmatrix}
    a_1  &   \bold{u}_1 \\
    \bold{v}_1 &   M^{(12)} 
\end{pmatrix} \,;
\qquad   M^{(2)}  = \begin{pmatrix}
    a_2  &   \bold{u}_2 \\
    \bold{v}_2 &   M^{(12)}
  \end{pmatrix} \,.
\ee
where~$\bold{u}_1$ and~$\bold{u}_2$ are $(N-2)$-dimensional row vectors, while~$\bold{v}_1$ and~$\bold{v}_2$ are column vectors of the same dimensionality.
The inverses are given by
\bea
\nonumber
& &   \left(M^{(1)}\right)^{-1}  = \begin{pmatrix}
     Q_1  & -Q_1 \bold{u}_1 W \\
     -W \bold{v}_1 Q_1 &  \big(\mathds{1}+ W\bold{v}_1Q_1 \bold{u}_1\big)W
  \end{pmatrix}  \,;\\ 
& &  \left(M^{(2)}\right)^{-1}  = \begin{pmatrix}
     Q_2 & -Q_2 \bold{u} W \\
     -W \bold{v}_2 Q_2 &  \big(\mathds{1} + W\bold{v}_2Q_2 \bold{u}_2 \big)W
  \end{pmatrix} 
 \label{eqn:Minverses}
\eea
where~$W\equiv\left(M^{(12)}\right)^{-1}$ and~$Q\equiv \left(a-\bold{u} W \bold{v}  \right)^{-1}$. From these inverses one can compute MFPTs using~\eqref{eqn:MFPTinversesum},
\bea
&&  T_{12} = \big(-1+\hat{e}_{N-2} W \bold{v}_1\big) Q_1,; \nonumber  \\ 
&& T_{21} = \big(-1+\hat{e}_{N-2} W \bold{v}_2\big) Q_2\,; \nonumber \\
&&  T_{1i} = \big(1-\hat{e}_{N-2} W \bold{v}_1\big) Q_1 \bold{u}_1 W - \hat{e}_{N-2} W = -T_{12}\bold{u}_1 W - \hat{e}_{N-2} W \,; \nonumber \\
 &&  T_{2i} =  \big(1 - \hat{e}_{N-2} W \bold{v}_2\big) Q_2 \bold{u}_2 W - \hat{e}_{N-2} W = -T_{21}\bold{u}_2 W - \hat{e}_{N-2} W \,.
\label{eqn:proof_sumrule1}
\eea
Meanwhile, since there is no leakage and no terminals, the sum of any column of $M$ vanishes, that is $M_{ii} =-\sum_{j\neq i} M_{ji}$. This condition can be expressed as
\be
   \bold{u}_1 + \bold{u}_2 + \hat{e}_{N-2} M^{(12)} = 0  \quad \Longrightarrow  \quad  \bold{u}_1 W + \bold{u}_2 W = -\hat{e}_{N-2}\,.
 \label{eqn:proof_sumrule2}
\ee

With~\eqref{eqn:proof_sumrule1} and~\eqref{eqn:proof_sumrule2} at hand we are ready to evaluate~$T_{21} T_{1i}+T_{12} T_{2i}$:
\bea
\nonumber
T_{1i}T_{21} + T_{2i}T_{12} & =& - T_{21}T_{12} \bold{u}_1 W - T_{12}T_{21} \bold{u}_2 W    - (T_{12}+T_{21}) \hat{e}_{N-2} W \\
& = &T_{21}T_{12} \hat{e}_{N-2} - (T_{12}+T_{21}) \hat{e}_{N-2} W\,.
\eea
Recalling that~$W=\left(M^{(12)}\right)^{-1}$, this is recognized as~\eqref{eqn:Trecur equiv}. Thus we have proven the relation~\eqref{eqn:Trecur}.  

\subsection{Three-target first-passage time}

For completeness, we note that by following the same logic, it is straightforward to show that the three-target search time~$T_{123,i}$ is given by: 
\begin{equation}
 \boxed{T_{123,i} = T_{12,i}+ T_{13,i}+ T_{23,i} -  T_{1i} - T_{2i} - T_{3i} - \left[\hat{e}_{N-3} \left(M^{(123)}\right)^{-1}\right]_i} \,,
 \end{equation}
where $M^{(123)}$ is the matrix without the first, second and third columns and rows. Furthermore, using the above techniques one can show that 
\be
\boxed{\left[\hat{e}_{N-3} \left(M^{(123)}\right)^{-1}\right]_i  = \frac{R_{13,2}R_{23,1}R_{12,i}+R_{12,3}R_{23,1}R_{13,i} +R_{13,2}R_{12,3}R_{23,i}- R_{13,2}R_{23,1}R_{12,3}}{R_{13,2}R_{23,1} + R_{12,3}R_{23,1}+ R_{13,2}R_{12,3}} } ,
\ee
where 
\begin{equation}
 R_{jk,i} = \left[\hat{e}_{N-3}\left(M^{(jk)}\right)^{-1} \right]_i = \frac{T_{jk}T_{kj} -T_{jk}T_{ki} - T_{kj}T_{ji} }{ T_{jk} +T_{kj}}.
\end{equation}
Therefore we see that the three target search time can be expressed solely in terms of single-target first-passage times, although the full expression is quite lengthy. 

This procedure can be generalized to compute the $n-$target search time. It is easy to deduce that the general $n-$target search time can be similarly expressed in terms of single-target first-passage times. 
The average time to visit every node in a region is known as the {\it cover time}~\cite{cover_time}.

\subsection{Escape time for a particular region}
\label{escape time proof}

Following the logic of the above section, we can derive an expression for the escape time from a particular subset of nodes~$\Omega$.
The probability that the random walker has never escaped~$\Omega$ at time~$t$, or survival probability, is given by
\begin{equation}
  P^{\rm S}_{j \in\Omega}(t) = \left[ \hat{e} \, e^{M_{\Omega}t} \right]_j\,.
\end{equation}
where~$M_{\Omega}$ is the submatrix of~$M$ that consists of rows and columns in~$\Omega$ only, while~$\hat{e}$ is the unit-entry vector of corresponding dimensionality. 
It is then straightforward to evaluate the escape time from any node~$j\in\Omega$:
\be
t^{\rm esc}_j = -\int_{0}^{\infty} \rd t \, t \frac{\rd}{\rd t}P^{\rm S}_{j\in\Omega}(t) = \left[-\hat{e} \, M_{\Omega}^{-1} \right]_j = -\sum_{i\in\Omega} M^{-1}_{ij} \,.
 \ee
This proves~\eqref{Tesc j}.

\subsection{Leaking region: ever hitting probability and first-passage statistics}
\label{apx:leaking}

Our next result is a derivation of the ever-hitting probability~${\cal P}^{\infty}_{kj}$ for some node~$k\in \Omega$, starting from any other node~$j\in \Omega$,
allowing for leakage out of the region~$\Omega$. Specifically, suppose that~$\Omega$ is in contact with an environment~${\cal S}$, such that there is leakage
from~$\Omega$ to~${\cal S}$ but no influx from~${\cal S}$ to~$\Omega$. In other words, a random walker starting in~$\Omega$ can escape the region but never re-enter.
The transition matrix from the whole system is then given by
\begin{equation}
{\cal M} = \begin{pmatrix}
   M_{\cal S} & L \\ 
   0 &  M
 \end{pmatrix}
\end{equation}
where~$M$ and~$M_{\cal S}$ represents transitions within~$\Omega$ and~${\cal S}$, respectively, while~$L$ encode the leakage from~$\Omega$ to~${\cal S}$. As a special case, if all nodes 
within~${\cal S}$ are terminals, then~$M_{\cal S}=0$. 

To derive the ever-hitting probability~${\cal P}^{\infty}_{kj}$, with $j,k \in \Omega$, note that 
\be
   \sum_{i}  \left[{\rm e}^{{\cal M}^{(k)}t } \right]_{ij}: \mbox{probability of not having visited node~$k\in \Omega$ by time~$t$}\,,
\ee
where, as before,~${\cal M}^{(k)}$ denotes the transition matrix without the~$k$-th row and column. The sum is over all nodes $\in\, \Omega\cup {\cal S}$,
which accounts for the fact that the walker may have stayed in~$\Omega$ but not reached~$k$ by time~$t$, or escaped to~${\cal S}$ some time earlier. 
Thus the ever-hitting probability is given by
\be
{\cal P}^{\infty}_{kj} = 1 - \lim_{t\to \infty} \sum_i \left[{\rm e}^{{\cal M}^{(k)} t} \right]_{ij}\,.
\label{Pever jk}
\ee
To evaluate this, note that 
\begin{equation}
\left({\cal M}^{(k)}\right)^n =  \begin{pmatrix}
    M_{\cal S} & L^{(k)} \\ 
   0 &  M^{(k)}
 \end{pmatrix}^n = 
 \begin{pmatrix}
    M_{\cal S}^n &  \sum_{m=1}^{n-1} M_{\cal S}^{n-1-m}L^{(k)} \left(M^{(k)}\right)^m  \\ 
   0 &  \left(M^{(k)}\right)^n
 \end{pmatrix}\,.
\end{equation}
It follows that, if~$j$ is a node in~$\Omega$, we have 
\be
 \sum_i \left[{\rm e}^{{\cal M}^{(k)} t} \right]_{ij} = \sum_{i \in {\cal S}}\left[ L^{(k)} \left(M^{(k)}\right)^{-1} \left({\rm e}^{M^{(k)} t} -\mathds{1} \right) \right]_{ij} + \sum_{i\in \Omega} \left[ {\rm e}^{M^{(k)} t} \right]_{ij}\,,
\ee
where we have used~$\sum_{i \in {\cal S}} M_{{\cal S} ij} =0$. Since~$M^{(k)}$ only has negative eigenvalues, as it represents a leaking region, we have~$\lim_{t\to \infty} {\rm e}^{M^{(k)} t} = 0$,  and hence
\be
\lim_{t\to \infty} \sum_i \left[{\rm e}^{{\cal M}^{(k)} t} \right]_{ij} = - \sum_{i }\left[ L^{(k)} \left(M^{(k)}\right)^{-1} \right]_{ij}\,.
\ee
Substituting into~\eqref{Pever jk} gives the desired result for the ever-hitting probability,
\be
 {\cal P}^{\infty}_{kj} = 1 +  \sum_{i }\left[ L^{(k)} \left(M^{(k)}\right)^{-1} \right]_{ij}  = - \sum_{\ell\neq k} M_{  k\ell} \left(M^{(k)}\right)^{-1}_{\ell j}\,,
 \ee
where in the last step we have used~$\sum_{a \in A} L^{(k)}_{a\ell} + \sum_{i \in \Omega}M_{ i\ell}^{(k)}  = -M_{ k\ell}$. 

\subsection{dS-only proxy region}
\label{F ineq appen}

Consider the proxy dS-only region discussed in Sec.~\ref{Kemeny constant}. In this Section we prove the inequality~\eqref{F ineq} relating first-passage densities with and without terminals. As in Sec.~\ref{Kemeny constant}, dS-only variables are denoted by hats.

First note that the dS-only first-passage density~$\hat{F}_{ij}$ satisfies, in the downward approximation, the dS-only equivalent of~\eqref{Fs}:
\be
\hat{F}_{ij}(s)\simeq \left(\mathds{1} - \hat{T}(s)\right)^{-1}_{ij}\,;\qquad \hat{T}_{ij}(s) = \frac{\kappa_{ij}}{s+ \hat{\kappa}_j}\,.
\ee
Notice that~$\kappa_{ij}$ is the same with or without terminals because it is a~${\rm dS}\rightarrow {\rm dS}$ transition rate, whereas~$\hat\kappa_j \neq \kappa_j$ because of decay channels into terminals. Indeed, since~$\hat{\kappa}_j\leq \kappa_j$, this implies
\be
\hat{T}_{ij}(s) \geq T_{ij}(s)\,,
\ee
and therefore
\be
F_{ij}(s) \leq \hat{F}_{ij}(s)\,.
\ee
By similar reasoning, it is straightforward to establish inequalities for all moments of the first-passage density, 
\be 
0\leq (-1)^n \left.\frac{{\rm d}^n F_{ij}(s)}{{\rm d}s^n} \right\vert_{s = 0} \leq (-1)^n \left.\frac{{\rm d}^n \hat{F}_{ij}(s)}{{\rm d}s^n} \right\vert_{s = 0}\,;\qquad n = 0,1,2,\ldots 
\ee
Therefore, the difference~$\hat{F}_{ij}(s) - F_{ij}(s)$ is a completely monotone function, and it follows that
\be
F_{ij}(t) \leq \hat{F}_{ij}(t)\,.
\ee
This proves~\eqref{F ineq}.

\bibliographystyle{utphys}
\bibliography{early_time_measure_submit.bib}

\providecommand{\href}[2]{#2}\begingroup\raggedright\begin{thebibliography}{100}

\bibitem{Frampton:1976kf}
P.~Frampton, ``{Vacuum Instability and Higgs Scalar Mass},''
  \href{http://dx.doi.org/10.1103/PhysRevLett.37.1378}{{\em Phys. Rev. Lett.}
  {\bfseries 37} (1976) 1378}. [Erratum: Phys.Rev.Lett. 37, 1716 (1976)].

\bibitem{Sher:1988mj}
M.~Sher, ``{Electroweak Higgs Potentials and Vacuum Stability},''
  \href{http://dx.doi.org/10.1016/0370-1573(89)90061-6}{{\em Phys. Rept.}
  {\bfseries 179} (1989) 273--418}.

\bibitem{Casas:1994qy}
J.~Casas, J.~Espinosa, and M.~Quiros, ``{Improved Higgs mass stability bound in
  the standard model and implications for supersymmetry},''
  \href{http://dx.doi.org/10.1016/0370-2693(94)01404-Z}{{\em Phys. Lett. B}
  {\bfseries 342} (1995) 171--179},
  \href{http://arxiv.org/abs/hep-ph/9409458}{{\ttfamily arXiv:hep-ph/9409458}}.

\bibitem{Espinosa:1995se}
J.~Espinosa and M.~Quiros, ``{Improved metastability bounds on the standard
  model Higgs mass},''
  \href{http://dx.doi.org/10.1016/0370-2693(95)00572-3}{{\em Phys. Lett. B}
  {\bfseries 353} (1995) 257--266},
  \href{http://arxiv.org/abs/hep-ph/9504241}{{\ttfamily arXiv:hep-ph/9504241}}.

\bibitem{Isidori:2001bm}
G.~Isidori, G.~Ridolfi, and A.~Strumia, ``{On the metastability of the standard
  model vacuum},'' \href{http://dx.doi.org/10.1016/S0550-3213(01)00302-9}{{\em
  Nucl. Phys. B} {\bfseries 609} (2001) 387--409},
  \href{http://arxiv.org/abs/hep-ph/0104016}{{\ttfamily arXiv:hep-ph/0104016}}.

\bibitem{Espinosa:2007qp}
J.~Espinosa, G.~Giudice, and A.~Riotto, ``{Cosmological implications of the
  Higgs mass measurement},''
  \href{http://dx.doi.org/10.1088/1475-7516/2008/05/002}{{\em JCAP} {\bfseries
  05} (2008) 002}, \href{http://arxiv.org/abs/0710.2484}{{\ttfamily
  arXiv:0710.2484 [hep-ph]}}.

\bibitem{Ellis:2009tp}
J.~Ellis, J.~Espinosa, G.~Giudice, A.~Hoecker, and A.~Riotto, ``{The Probable
  Fate of the Standard Model},''
  \href{http://dx.doi.org/10.1016/j.physletb.2009.07.054}{{\em Phys. Lett. B}
  {\bfseries 679} (2009) 369--375},
  \href{http://arxiv.org/abs/0906.0954}{{\ttfamily arXiv:0906.0954 [hep-ph]}}.

\bibitem{Degrassi:2012ry}
G.~Degrassi, S.~Di~Vita, J.~Elias-Miro, J.~R. Espinosa, G.~F. Giudice,
  G.~Isidori, and A.~Strumia, ``{Higgs mass and vacuum stability in the
  Standard Model at NNLO},''
  \href{http://dx.doi.org/10.1007/JHEP08(2012)098}{{\em JHEP} {\bfseries 08}
  (2012) 098}, \href{http://arxiv.org/abs/1205.6497}{{\ttfamily arXiv:1205.6497
  [hep-ph]}}.

\bibitem{Buttazzo:2013uya}
D.~Buttazzo, G.~Degrassi, P.~P. Giardino, G.~F. Giudice, F.~Sala, A.~Salvio,
  and A.~Strumia, ``{Investigating the near-criticality of the Higgs boson},''
  \href{http://dx.doi.org/10.1007/JHEP12(2013)089}{{\em JHEP} {\bfseries 12}
  (2013) 089}, \href{http://arxiv.org/abs/1307.3536}{{\ttfamily arXiv:1307.3536
  [hep-ph]}}.

\bibitem{Lalak:2014qua}
Z.~Lalak, M.~Lewicki, and P.~Olszewski, ``{Higher-order scalar interactions and
  SM vacuum stability},'' \href{http://dx.doi.org/10.1007/JHEP05(2014)119}{{\em
  JHEP} {\bfseries 05} (2014) 119},
  \href{http://arxiv.org/abs/1402.3826}{{\ttfamily arXiv:1402.3826 [hep-ph]}}.

\bibitem{Andreassen:2014gha}
A.~Andreassen, W.~Frost, and M.~D. Schwartz, ``{Consistent Use of the Standard
  Model Effective Potential},''
  \href{http://dx.doi.org/10.1103/PhysRevLett.113.241801}{{\em Phys. Rev.
  Lett.} {\bfseries 113} no.~24, (2014) 241801},
  \href{http://arxiv.org/abs/1408.0292}{{\ttfamily arXiv:1408.0292 [hep-ph]}}.

\bibitem{Branchina:2014rva}
V.~Branchina, E.~Messina, and M.~Sher, ``{Lifetime of the electroweak vacuum
  and sensitivity to Planck scale physics},''
  \href{http://dx.doi.org/10.1103/PhysRevD.91.013003}{{\em Phys. Rev. D}
  {\bfseries 91} (2015) 013003},
  \href{http://arxiv.org/abs/1408.5302}{{\ttfamily arXiv:1408.5302 [hep-ph]}}.

\bibitem{Bednyakov:2015sca}
A.~Bednyakov, B.~Kniehl, A.~Pikelner, and O.~Veretin, ``{Stability of the
  Electroweak Vacuum: Gauge Independence and Advanced Precision},''
  \href{http://dx.doi.org/10.1103/PhysRevLett.115.201802}{{\em Phys. Rev.
  Lett.} {\bfseries 115} no.~20, (2015) 201802},
  \href{http://arxiv.org/abs/1507.08833}{{\ttfamily arXiv:1507.08833
  [hep-ph]}}.

\bibitem{Iacobellis:2016eof}
G.~Iacobellis and I.~Masina, ``{Stationary configurations of the Standard Model
  Higgs potential: electroweak stability and rising inflection point},''
  \href{http://dx.doi.org/10.1103/PhysRevD.94.073005}{{\em Phys. Rev. D}
  {\bfseries 94} no.~7, (2016) 073005},
  \href{http://arxiv.org/abs/1604.06046}{{\ttfamily arXiv:1604.06046
  [hep-ph]}}.

\bibitem{Andreassen:2017rzq}
A.~Andreassen, W.~Frost, and M.~D. Schwartz, ``{Scale Invariant Instantons and
  the Complete Lifetime of the Standard Model},''
  \href{http://dx.doi.org/10.1103/PhysRevD.97.056006}{{\em Phys. Rev. D}
  {\bfseries 97} no.~5, (2018) 056006},
  \href{http://arxiv.org/abs/1707.08124}{{\ttfamily arXiv:1707.08124
  [hep-ph]}}.

\bibitem{Giudice:2006sn}
G.~Giudice and R.~Rattazzi, ``{Living Dangerously with Low-Energy
  Supersymmetry},''
  \href{http://dx.doi.org/10.1016/j.nuclphysb.2006.07.031}{{\em Nucl. Phys. B}
  {\bfseries 757} (2006) 19--46},
  \href{http://arxiv.org/abs/hep-ph/0606105}{{\ttfamily arXiv:hep-ph/0606105}}.

\bibitem{Friedrich:1987}
H.~Friedrich, ``{On the existence of n-geodesically complete or future complete
  solutions of Einstein's field equations with smooth asymptotic structure},''
  {\em Communications in Mathematical Physics} {\bfseries 107} no.~4, (Dec.,
  1986) 587--609.

\bibitem{Bizon:2011gg}
P.~Bizon and A.~Rostworowski, ``{On weakly turbulent instability of anti-de
  Sitter space},'' \href{http://dx.doi.org/10.1103/PhysRevLett.107.031102}{{\em
  Phys. Rev. Lett.} {\bfseries 107} (2011) 031102},
  \href{http://arxiv.org/abs/1104.3702}{{\ttfamily arXiv:1104.3702 [gr-qc]}}.

\bibitem{Strominger:2001pn}
A.~Strominger, ``{The dS / CFT correspondence},''
  \href{http://dx.doi.org/10.1088/1126-6708/2001/10/034}{{\em JHEP} {\bfseries
  10} (2001) 034}, \href{http://arxiv.org/abs/hep-th/0106113}{{\ttfamily
  arXiv:hep-th/0106113}}.

\bibitem{Steinhardt:1982kg}
P.~J. Steinhardt, ``{NATURAL INFLATION},'' in {\em {Nuffield Workshop on the
  Very Early Universe}}, pp.~251--266.
\newblock 7, 1982.

\bibitem{Vilenkin:1983xq}
A.~Vilenkin, ``{The Birth of Inflationary Universes},''
  \href{http://dx.doi.org/10.1103/PhysRevD.27.2848}{{\em Phys. Rev. D}
  {\bfseries 27} (1983) 2848}.

\bibitem{Linde:1986fc}
A.~D. Linde, ``{Eternal Chaotic Inflation},''
  \href{http://dx.doi.org/10.1142/S0217732386000129}{{\em Mod. Phys. Lett. A}
  {\bfseries 1} (1986) 81}.

\bibitem{Linde:1986fd}
A.~D. Linde, ``{Eternally Existing Selfreproducing Chaotic Inflationary
  Universe},'' \href{http://dx.doi.org/10.1016/0370-2693(86)90611-8}{{\em Phys.
  Lett. B} {\bfseries 175} (1986) 395--400}.

\bibitem{Starobinsky:1986fx}
A.~A. Starobinsky, ``{Stochastic de Sitter (Inflationary) Stage in the Early
  Universe},'' \href{http://dx.doi.org/10.1007/3-540-16452-9_6}{{\em Lect.
  Notes Phys.} {\bfseries 246} (1986) 107--126}.

\bibitem{Obied:2018sgi}
G.~Obied, H.~Ooguri, L.~Spodyneiko, and C.~Vafa, ``{De Sitter Space and the
  Swampland},'' \href{http://arxiv.org/abs/1806.08362}{{\ttfamily
  arXiv:1806.08362 [hep-th]}}.

\bibitem{Agrawal:2018own}
P.~Agrawal, G.~Obied, P.~J. Steinhardt, and C.~Vafa, ``{On the Cosmological
  Implications of the String Swampland},''
  \href{http://dx.doi.org/10.1016/j.physletb.2018.07.040}{{\em Phys. Lett. B}
  {\bfseries 784} (2018) 271--276},
  \href{http://arxiv.org/abs/1806.09718}{{\ttfamily arXiv:1806.09718
  [hep-th]}}.

\bibitem{Guth:2007ng}
A.~H. Guth, ``{Eternal inflation and its implications},''
  \href{http://dx.doi.org/10.1088/1751-8113/40/25/S25}{{\em J. Phys. A}
  {\bfseries 40} (2007) 6811--6826},
  \href{http://arxiv.org/abs/hep-th/0702178}{{\ttfamily arXiv:hep-th/0702178}}.

\bibitem{Freivogel:2011eg}
B.~Freivogel, ``{Making predictions in the multiverse},''
  \href{http://dx.doi.org/10.1088/0264-9381/28/20/204007}{{\em Class. Quant.
  Grav.} {\bfseries 28} (2011) 204007},
  \href{http://arxiv.org/abs/1105.0244}{{\ttfamily arXiv:1105.0244 [hep-th]}}.

\bibitem{Linde:1993nz}
A.~D. Linde and A.~Mezhlumian, ``{Stationary universe},''
  \href{http://dx.doi.org/10.1016/0370-2693(93)90187-M}{{\em Phys. Lett. B}
  {\bfseries 307} (1993) 25--33},
  \href{http://arxiv.org/abs/gr-qc/9304015}{{\ttfamily arXiv:gr-qc/9304015}}.

\bibitem{Linde:1993xx}
A.~D. Linde, D.~A. Linde, and A.~Mezhlumian, ``{From the Big Bang theory to the
  theory of a stationary universe},''
  \href{http://dx.doi.org/10.1103/PhysRevD.49.1783}{{\em Phys. Rev. D}
  {\bfseries 49} (1994) 1783--1826},
  \href{http://arxiv.org/abs/gr-qc/9306035}{{\ttfamily arXiv:gr-qc/9306035}}.

\bibitem{GarciaBellido:1993wn}
J.~Garcia-Bellido, A.~D. Linde, and D.~A. Linde, ``{Fluctuations of the
  gravitational constant in the inflationary Brans-Dicke cosmology},''
  \href{http://dx.doi.org/10.1103/PhysRevD.50.730}{{\em Phys. Rev. D}
  {\bfseries 50} (1994) 730--750},
  \href{http://arxiv.org/abs/astro-ph/9312039}{{\ttfamily
  arXiv:astro-ph/9312039}}.

\bibitem{Vilenkin:1994ua}
A.~Vilenkin, ``{Predictions from quantum cosmology},''
  \href{http://dx.doi.org/10.1103/PhysRevLett.74.846}{{\em Phys. Rev. Lett.}
  {\bfseries 74} (1995) 846--849},
  \href{http://arxiv.org/abs/gr-qc/9406010}{{\ttfamily arXiv:gr-qc/9406010}}.

\bibitem{Garriga:1997ef}
J.~Garriga and A.~Vilenkin, ``{Recycling universe},''
  \href{http://dx.doi.org/10.1103/PhysRevD.57.2230}{{\em Phys. Rev. D}
  {\bfseries 57} (1998) 2230--2244},
  \href{http://arxiv.org/abs/astro-ph/9707292}{{\ttfamily
  arXiv:astro-ph/9707292}}.

\bibitem{Garriga:2005av}
J.~Garriga, D.~Schwartz-Perlov, A.~Vilenkin, and S.~Winitzki, ``{Probabilities
  in the inflationary multiverse},''
  \href{http://dx.doi.org/10.1088/1475-7516/2006/01/017}{{\em JCAP} {\bfseries
  01} (2006) 017}, \href{http://arxiv.org/abs/hep-th/0509184}{{\ttfamily
  arXiv:hep-th/0509184}}.

\bibitem{Albrecht:2002uz}
A.~Albrecht, ``{Cosmic inflation and the arrow of time},''
  \href{http://arxiv.org/abs/astro-ph/0210527}{{\ttfamily
  arXiv:astro-ph/0210527}}.

\bibitem{Dyson:2002pf}
L.~Dyson, M.~Kleban, and L.~Susskind, ``{Disturbing implications of a
  cosmological constant},''
  \href{http://dx.doi.org/10.1088/1126-6708/2002/10/011}{{\em JHEP} {\bfseries
  10} (2002) 011}, \href{http://arxiv.org/abs/hep-th/0208013}{{\ttfamily
  arXiv:hep-th/0208013}}.

\bibitem{Albrecht:2004ke}
A.~Albrecht and L.~Sorbo, ``{Can the universe afford inflation?},''
  \href{http://dx.doi.org/10.1103/PhysRevD.70.063528}{{\em Phys. Rev. D}
  {\bfseries 70} (2004) 063528},
  \href{http://arxiv.org/abs/hep-th/0405270}{{\ttfamily arXiv:hep-th/0405270}}.

\bibitem{Page:2005ur}
D.~N. Page, ``{The Lifetime of the universe},'' {\em J. Korean Phys. Soc.}
  {\bfseries 49} (2006) 711--714,
  \href{http://arxiv.org/abs/hep-th/0510003}{{\ttfamily arXiv:hep-th/0510003}}.

\bibitem{Page:2006dt}
D.~N. Page, ``{Is our universe likely to decay within 20 billion years?},''
  \href{http://dx.doi.org/10.1103/PhysRevD.78.063535}{{\em Phys. Rev. D}
  {\bfseries 78} (2008) 063535},
  \href{http://arxiv.org/abs/hep-th/0610079}{{\ttfamily arXiv:hep-th/0610079}}.

\bibitem{DeSimone:2008if}
A.~De~Simone, A.~H. Guth, A.~D. Linde, M.~Noorbala, M.~P. Salem, and
  A.~Vilenkin, ``{Boltzmann brains and the scale-factor cutoff measure of the
  multiverse},'' \href{http://dx.doi.org/10.1103/PhysRevD.82.063520}{{\em Phys.
  Rev. D} {\bfseries 82} (2010) 063520},
  \href{http://arxiv.org/abs/0808.3778}{{\ttfamily arXiv:0808.3778 [hep-th]}}.

\bibitem{Bousso:2008hz}
R.~Bousso, B.~Freivogel, and I.-S. Yang, ``{Properties of the scale factor
  measure},'' \href{http://dx.doi.org/10.1103/PhysRevD.79.063513}{{\em Phys.
  Rev. D} {\bfseries 79} (2009) 063513},
  \href{http://arxiv.org/abs/0808.3770}{{\ttfamily arXiv:0808.3770 [hep-th]}}.

\bibitem{Hartle:2007zv}
J.~B. Hartle and M.~Srednicki, ``{Are we typical?},''
  \href{http://dx.doi.org/10.1103/PhysRevD.75.123523}{{\em Phys. Rev. D}
  {\bfseries 75} (2007) 123523},
  \href{http://arxiv.org/abs/0704.2630}{{\ttfamily arXiv:0704.2630 [hep-th]}}.

\bibitem{Srednicki:2009vb}
M.~Srednicki and J.~Hartle, ``{Science in a Very Large Universe},''
  \href{http://dx.doi.org/10.1103/PhysRevD.81.123524}{{\em Phys. Rev. D}
  {\bfseries 81} (2010) 123524},
  \href{http://arxiv.org/abs/0906.0042}{{\ttfamily arXiv:0906.0042 [hep-th]}}.

\bibitem{Khoury:2019yoo}
J.~Khoury and O.~Parrikar, ``{Search Optimization, Funnel Topography, and
  Dynamical Criticality on the String Landscape},''
  \href{http://dx.doi.org/10.1088/1475-7516/2019/12/014}{{\em JCAP} {\bfseries
  12} (2019) 014}, \href{http://arxiv.org/abs/1907.07693}{{\ttfamily
  arXiv:1907.07693 [hep-th]}}.

\bibitem{Khoury:2019ajl}
J.~Khoury, ``{Accessibility Measure for Eternal Inflation: Dynamical
  Criticality and Higgs Metastability},''
  \href{http://dx.doi.org/10.1088/1475-7516/2021/06/009}{{\em JCAP} {\bfseries
  06} (2021) 009}, \href{http://arxiv.org/abs/1912.06706}{{\ttfamily
  arXiv:1912.06706 [hep-th]}}.

\bibitem{Kartvelishvili:2020thd}
G.~Kartvelishvili, J.~Khoury, and A.~Sharma, ``{The Self-Organized Critical
  Multiverse},'' \href{http://dx.doi.org/10.1088/1475-7516/2021/02/028}{{\em
  JCAP} {\bfseries 02} (2021) 028},
  \href{http://arxiv.org/abs/2003.12594}{{\ttfamily arXiv:2003.12594
  [hep-th]}}.

\bibitem{Garriga:2007wz}
J.~Garriga and A.~Vilenkin, ``{Prediction and explanation in the multiverse},''
  \href{http://dx.doi.org/10.1103/PhysRevD.77.043526}{{\em Phys. Rev. D}
  {\bfseries 77} (2008) 043526},
  \href{http://arxiv.org/abs/0711.2559}{{\ttfamily arXiv:0711.2559 [hep-th]}}.

\bibitem{Denef:2017cxt}
F.~Denef, M.~R. Douglas, B.~Greene, and C.~Zukowski, ``{Computational
  complexity of the landscape II\textemdash{}Cosmological considerations},''
  \href{http://dx.doi.org/10.1016/j.aop.2018.03.013}{{\em Annals Phys.}
  {\bfseries 392} (2018) 93--127},
  \href{http://arxiv.org/abs/1706.06430}{{\ttfamily arXiv:1706.06430
  [hep-th]}}.

\bibitem{closeness1}
A.~Bavelas, ``{Communication Patterns in Task-Oriented Groups},''
  \href{http://dx.doi.org/10.1121/1.1906679}{{\em Acoustical Society of America
  Journal} {\bfseries 22} no.~6, (Jan., 1950) 725}.

\bibitem{closeness2}
G.~Sabidussi, ``The centrality index of a graph,'' {\em Psychometrika}
  {\bfseries 31} no.~4, (1966) 581--603.

\bibitem{Giudice:2021viw}
G.~F. Giudice, M.~McCullough, and T.~You, ``{Self-Organised Localisation},''
  \href{http://arxiv.org/abs/2105.08617}{{\ttfamily arXiv:2105.08617
  [hep-ph]}}.

\bibitem{SchwartzPerlov:2006hi}
D.~Schwartz-Perlov and A.~Vilenkin, ``{Probabilities in the Bousso-Polchinski
  multiverse},'' \href{http://dx.doi.org/10.1088/1475-7516/2006/06/010}{{\em
  JCAP} {\bfseries 06} (2006) 010},
  \href{http://arxiv.org/abs/hep-th/0601162}{{\ttfamily arXiv:hep-th/0601162}}.

\bibitem{Olum:2007yk}
K.~D. Olum and D.~Schwartz-Perlov, ``{Anthropic prediction in a large toy
  landscape},'' \href{http://dx.doi.org/10.1088/1475-7516/2007/10/010}{{\em
  JCAP} {\bfseries 10} (2007) 010},
  \href{http://arxiv.org/abs/0705.2562}{{\ttfamily arXiv:0705.2562 [hep-th]}}.
  [Erratum: JCAP 10, E02 (2019)].

\bibitem{Lee:1987qc}
K.-M. Lee and E.~J. Weinberg, ``{Decay of the True Vacuum in Curved
  Space-time},'' \href{http://dx.doi.org/10.1103/PhysRevD.36.1088}{{\em Phys.
  Rev. D} {\bfseries 36} (1987) 1088}.

\bibitem{Bousso:2006ev}
R.~Bousso, ``{Holographic probabilities in eternal inflation},''
  \href{http://dx.doi.org/10.1103/PhysRevLett.97.191302}{{\em Phys. Rev. Lett.}
  {\bfseries 97} (2006) 191302},
  \href{http://arxiv.org/abs/hep-th/0605263}{{\ttfamily arXiv:hep-th/0605263}}.

\bibitem{Redner}
S.~Redner, {\em A guide to first-passage processes}.
\newblock Cambridge University Press, Cambridge, 2001.

\bibitem{proteins1}
J.~D. {Bryngelson}, J.~N. {Onuchic}, N.~D. {Socci}, and P.~G. {Wolynes},
  ``{Funnels, Pathways and the Energy Landscape of Protein Folding: A
  Synthesis},'' {\em Proteins-Struct. Func. and Genetics} {\bfseries 21} (1995)
  167, \href{http://arxiv.org/abs/chem-ph/9411008}{{\ttfamily
  arXiv:chem-ph/9411008}}.

\bibitem{kemeny}
J.~Kemeny and J.~Snell, {\em Finite markov chains}.
\newblock van Nostrand Princeton, NJ, 1960.

\bibitem{compPT1}
R.~{Monasson}, R.~{Zecchina}, S.~{Kirkpatrick}, B.~{Selman}, and
  L.~{Troyansky}, ``{Determining computational complexity from characteristic
  `phase transitions'},'' {\em Nature} {\bfseries 400} no.~6740, (July, 1999)
  133--137.

\bibitem{compPT2}
T.~Hogg, B.~A. Huberman, and C.~P. Williams, ``Phase transitions and the search
  problem,'' {\em Artificial Intelligence} {\bfseries 81} (1996) 1.

\bibitem{Denef:2006ad}
F.~Denef and M.~R. Douglas, ``{Computational complexity of the landscape.
  I.},'' \href{http://dx.doi.org/10.1016/j.aop.2006.07.013}{{\em Annals Phys.}
  {\bfseries 322} (2007) 1096--1142},
  \href{http://arxiv.org/abs/hep-th/0602072}{{\ttfamily arXiv:hep-th/0602072}}.

\bibitem{Halverson:2018cio}
J.~Halverson and F.~Ruehle, ``{Computational Complexity of Vacua and Near-Vacua
  in Field and String Theory},''
  \href{http://dx.doi.org/10.1103/PhysRevD.99.046015}{{\em Phys. Rev. D}
  {\bfseries 99} no.~4, (2019) 046015},
  \href{http://arxiv.org/abs/1809.08279}{{\ttfamily arXiv:1809.08279
  [hep-th]}}.

\bibitem{Bao:2017thx}
N.~Bao, R.~Bousso, S.~Jordan, and B.~Lackey, ``{Fast optimization algorithms
  and the cosmological constant},''
  \href{http://dx.doi.org/10.1103/PhysRevD.96.103512}{{\em Phys. Rev. D}
  {\bfseries 96} no.~10, (2017) 103512},
  \href{http://arxiv.org/abs/1706.08503}{{\ttfamily arXiv:1706.08503
  [hep-th]}}.

\bibitem{thomas}
T.~Steingasser and J.~Khoury, ``{Particle physics implications of Page-time
  Higgs metastability},'' {\em to appear} .

\bibitem{Bousso:2000xa}
R.~Bousso and J.~Polchinski, ``{Quantization of four form fluxes and dynamical
  neutralization of the cosmological constant},''
  \href{http://dx.doi.org/10.1088/1126-6708/2000/06/006}{{\em JHEP} {\bfseries
  06} (2000) 006}, \href{http://arxiv.org/abs/hep-th/0004134}{{\ttfamily
  arXiv:hep-th/0004134}}.

\bibitem{Carifio:2017nyb}
J.~Carifio, W.~J. Cunningham, J.~Halverson, D.~Krioukov, C.~Long, and B.~D.
  Nelson, ``{Vacuum Selection from Cosmology on Networks of String
  Geometries},'' \href{http://dx.doi.org/10.1103/PhysRevLett.121.101602}{{\em
  Phys. Rev. Lett.} {\bfseries 121} no.~10, (2018) 101602},
  \href{http://arxiv.org/abs/1711.06685}{{\ttfamily arXiv:1711.06685
  [hep-th]}}.

\bibitem{Coleman:1977py}
S.~R. Coleman, ``{The Fate of the False Vacuum. 1. Semiclassical Theory},''
  \href{http://dx.doi.org/10.1103/PhysRevD.16.1248}{{\em Phys. Rev. D}
  {\bfseries 15} (1977) 2929--2936}. [Erratum: Phys.Rev.D 16, 1248 (1977)].

\bibitem{Callan:1977pt}
J.~Callan, Curtis~G. and S.~R. Coleman, ``{The Fate of the False Vacuum. 2.
  First Quantum Corrections},''
  \href{http://dx.doi.org/10.1103/PhysRevD.16.1762}{{\em Phys. Rev. D}
  {\bfseries 16} (1977) 1762--1768}.

\bibitem{Coleman:1980aw}
S.~R. Coleman and F.~De~Luccia, ``{Gravitational Effects on and of Vacuum
  Decay},'' \href{http://dx.doi.org/10.1103/PhysRevD.21.3305}{{\em Phys. Rev.
  D} {\bfseries 21} (1980) 3305}.

\bibitem{deAlwis:2019dkc}
S.~P. De~Alwis, F.~Muia, V.~Pasquarella, and F.~Quevedo, ``{Quantum Transitions
  Between Minkowski and de Sitter Spacetimes},''
  \href{http://dx.doi.org/10.1002/prop.202000069}{{\em Fortsch. Phys.}
  {\bfseries 68} no.~9, (2020) 2000069},
  \href{http://arxiv.org/abs/1909.01975}{{\ttfamily arXiv:1909.01975
  [hep-th]}}.

\bibitem{Cespedes:2020xpn}
S.~Cespedes, S.~P. de~Alwis, F.~Muia, and F.~Quevedo, ``{Lorentzian Vacuum
  Transitions: Open or Closed Universes?},''
  \href{http://arxiv.org/abs/2011.13936}{{\ttfamily arXiv:2011.13936
  [hep-th]}}.

\bibitem{Finelli:2008zg}
F.~Finelli, G.~Marozzi, A.~A. Starobinsky, G.~P. Vacca, and G.~Venturi,
  ``{Generation of fluctuations during inflation: Comparison of stochastic and
  field-theoretic approaches},''
  \href{http://dx.doi.org/10.1103/PhysRevD.79.044007}{{\em Phys. Rev. D}
  {\bfseries 79} (2009) 044007},
  \href{http://arxiv.org/abs/0808.1786}{{\ttfamily arXiv:0808.1786 [hep-th]}}.

\bibitem{Finelli:2010sh}
F.~Finelli, G.~Marozzi, A.~A. Starobinsky, G.~P. Vacca, and G.~Venturi,
  ``{Stochastic growth of quantum fluctuations during slow-roll inflation},''
  \href{http://dx.doi.org/10.1103/PhysRevD.82.064020}{{\em Phys. Rev. D}
  {\bfseries 82} (2010) 064020},
  \href{http://arxiv.org/abs/1003.1327}{{\ttfamily arXiv:1003.1327 [hep-th]}}.

\bibitem{Vennin:2015hra}
V.~Vennin and A.~A. Starobinsky, ``{Correlation Functions in Stochastic
  Inflation},'' \href{http://dx.doi.org/10.1140/epjc/s10052-015-3643-y}{{\em
  Eur. Phys. J. C} {\bfseries 75} (2015) 413},
  \href{http://arxiv.org/abs/1506.04732}{{\ttfamily arXiv:1506.04732
  [hep-th]}}.

\bibitem{Hawking:1981fz}
S.~W. Hawking and I.~G. Moss, ``{Supercooled Phase Transitions in the Very
  Early Universe},'' \href{http://dx.doi.org/10.1016/0370-2693(82)90946-7}{{\em
  Phys. Lett. B} {\bfseries 110} (1982) 35--38}.

\bibitem{Brown:1988kg}
J.~D. Brown and C.~Teitelboim, ``{Neutralization of the Cosmological Constant
  by Membrane Creation},''
  \href{http://dx.doi.org/10.1016/0550-3213(88)90559-7}{{\em Nucl. Phys. B}
  {\bfseries 297} (1988) 787--836}.

\bibitem{Assadullahi:2016gkk}
H.~Assadullahi, H.~Firouzjahi, M.~Noorbala, V.~Vennin, and D.~Wands,
  ``{Multiple Fields in Stochastic Inflation},''
  \href{http://dx.doi.org/10.1088/1475-7516/2016/06/043}{{\em JCAP} {\bfseries
  06} (2016) 043}, \href{http://arxiv.org/abs/1604.04502}{{\ttfamily
  arXiv:1604.04502 [hep-th]}}.

\bibitem{Vennin:2016wnk}
V.~Vennin, H.~Assadullahi, H.~Firouzjahi, M.~Noorbala, and D.~Wands,
  ``{Critical Number of Fields in Stochastic Inflation},''
  \href{http://dx.doi.org/10.1103/PhysRevLett.118.031301}{{\em Phys. Rev.
  Lett.} {\bfseries 118} no.~3, (2017) 031301},
  \href{http://arxiv.org/abs/1604.06017}{{\ttfamily arXiv:1604.06017
  [astro-ph.CO]}}.

\bibitem{Noorbala:2018zlv}
M.~Noorbala, V.~Vennin, H.~Assadullahi, H.~Firouzjahi, and D.~Wands,
  ``{Tunneling in Stochastic Inflation},''
  \href{http://dx.doi.org/10.1088/1475-7516/2018/09/032}{{\em JCAP} {\bfseries
  09} (2018) 032}, \href{http://arxiv.org/abs/1806.09634}{{\ttfamily
  arXiv:1806.09634 [hep-th]}}.

\bibitem{OU}
C.~Gardiner, {\em Handbook of Stochastic Methods: for Physics, Chemistry and
  the Natural Sciences}.
\newblock Springer Series in Synergetics (Book 13), Springer, 2004.

\bibitem{glassylandscape}
J.~C. Mauro and M.~M. Smedskjaer, ``{Statistical mechanics of glass},'' {\em
  Journal of Non-Crystalline Solids} {\bfseries 396} (2014) 41.

\bibitem{Weinberg:2000qm}
S.~Weinberg, ``{A Priori probability distribution of the cosmological
  constant},'' \href{http://dx.doi.org/10.1103/PhysRevD.61.103505}{{\em Phys.
  Rev. D} {\bfseries 61} (2000) 103505},
  \href{http://arxiv.org/abs/astro-ph/0002387}{{\ttfamily
  arXiv:astro-ph/0002387}}.

\bibitem{disorderedmedia}
S.~Havlin and D.~Ben-Avraham, ``Diffusion in disordered media,'' {\em Advances
  in Physics} {\bfseries 36} (1987) 695.

\bibitem{Page:1993wv}
D.~N. Page, ``{Information in black hole radiation},''
  \href{http://dx.doi.org/10.1103/PhysRevLett.71.3743}{{\em Phys. Rev. Lett.}
  {\bfseries 71} (1993) 3743--3746},
  \href{http://arxiv.org/abs/hep-th/9306083}{{\ttfamily arXiv:hep-th/9306083}}.

\bibitem{Danielsson:2002td}
U.~H. Danielsson, D.~Domert, and M.~E. Olsson, ``{Miracles and complementarity
  in de Sitter space},''
  \href{http://dx.doi.org/10.1103/PhysRevD.68.083508}{{\em Phys. Rev. D}
  {\bfseries 68} (2003) 083508},
  \href{http://arxiv.org/abs/hep-th/0210198}{{\ttfamily arXiv:hep-th/0210198}}.

\bibitem{Danielsson:2003wb}
U.~H. Danielsson and M.~E. Olsson, ``{On thermalization in de Sitter space},''
  \href{http://dx.doi.org/10.1088/1126-6708/2004/03/036}{{\em JHEP} {\bfseries
  03} (2004) 036}, \href{http://arxiv.org/abs/hep-th/0309163}{{\ttfamily
  arXiv:hep-th/0309163}}.

\bibitem{Ferreira:2016hee}
R.~Z. Ferreira, M.~Sandora, and M.~S. Sloth, ``{Asymptotic Symmetries in de
  Sitter and Inflationary Spacetimes},''
  \href{http://dx.doi.org/10.1088/1475-7516/2017/04/033}{{\em JCAP} {\bfseries
  04} (2017) 033}, \href{http://arxiv.org/abs/1609.06318}{{\ttfamily
  arXiv:1609.06318 [hep-th]}}.

\bibitem{Ferreira:2017ogo}
R.~Z. Ferreira, M.~Sandora, and M.~S. Sloth, ``{Patient Observers and
  Non-perturbative Infrared Dynamics in Inflation},''
  \href{http://dx.doi.org/10.1088/1475-7516/2018/02/055}{{\em JCAP} {\bfseries
  02} (2018) 055}, \href{http://arxiv.org/abs/1703.10162}{{\ttfamily
  arXiv:1703.10162 [hep-th]}}.

\bibitem{Creminelli:2008es}
P.~Creminelli, S.~Dubovsky, A.~Nicolis, L.~Senatore, and M.~Zaldarriaga, ``{The
  Phase Transition to Slow-roll Eternal Inflation},''
  \href{http://dx.doi.org/10.1088/1126-6708/2008/09/036}{{\em JHEP} {\bfseries
  09} (2008) 036}, \href{http://arxiv.org/abs/0802.1067}{{\ttfamily
  arXiv:0802.1067 [hep-th]}}.

\bibitem{Dubovsky:2008rf}
S.~Dubovsky, L.~Senatore, and G.~Villadoro, ``{The Volume of the Universe after
  Inflation and de Sitter Entropy},''
  \href{http://dx.doi.org/10.1088/1126-6708/2009/04/118}{{\em JHEP} {\bfseries
  04} (2009) 118}, \href{http://arxiv.org/abs/0812.2246}{{\ttfamily
  arXiv:0812.2246 [hep-th]}}.

\bibitem{Dubovsky:2011uy}
S.~Dubovsky, L.~Senatore, and G.~Villadoro, ``{Universality of the Volume Bound
  in Slow-Roll Eternal Inflation},''
  \href{http://dx.doi.org/10.1007/JHEP05(2012)035}{{\em JHEP} {\bfseries 05}
  (2012) 035}, \href{http://arxiv.org/abs/1111.1725}{{\ttfamily arXiv:1111.1725
  [hep-th]}}.

\bibitem{ArkaniHamed:2007ky}
N.~Arkani-Hamed, S.~Dubovsky, A.~Nicolis, E.~Trincherini, and G.~Villadoro,
  ``{A Measure of de Sitter entropy and eternal inflation},'' {\em JHEP}
  {\bfseries 05} (2007) 055, \href{http://arxiv.org/abs/0704.1814}{{\ttfamily
  arXiv:0704.1814 [hep-th]}}.

\bibitem{Sumitomo:2012wa}
Y.~Sumitomo and S.-H. Tye, ``{A Stringy Mechanism for A Small Cosmological
  Constant},'' \href{http://dx.doi.org/10.1088/1475-7516/2012/08/032}{{\em
  JCAP} {\bfseries 08} (2012) 032},
  \href{http://arxiv.org/abs/1204.5177}{{\ttfamily arXiv:1204.5177 [hep-th]}}.

\bibitem{Sumitomo:2012vx}
Y.~Sumitomo and S.~H. Tye, ``{A Stringy Mechanism for A Small Cosmological
  Constant - Multi-Moduli Cases -},''
  \href{http://dx.doi.org/10.1088/1475-7516/2013/02/006}{{\em JCAP} {\bfseries
  02} (2013) 006}, \href{http://arxiv.org/abs/1209.5086}{{\ttfamily
  arXiv:1209.5086 [hep-th]}}.

\bibitem{Sumitomo:2012cf}
Y.~Sumitomo and S.-H. Tye, ``{Preference for a Vanishingly Small Cosmological
  Constant in Supersymmetric Vacua in a Type IIB String Theory Model},''
  \href{http://dx.doi.org/10.1016/j.physletb.2013.05.027}{{\em Phys. Lett. B}
  {\bfseries 723} (2013) 406--410},
  \href{http://arxiv.org/abs/1211.6858}{{\ttfamily arXiv:1211.6858 [hep-th]}}.

\bibitem{Danielsson:2012by}
U.~Danielsson and G.~Dibitetto, ``{On the distribution of stable de Sitter
  vacua},'' \href{http://dx.doi.org/10.1007/JHEP03(2013)018}{{\em JHEP}
  {\bfseries 03} (2013) 018}, \href{http://arxiv.org/abs/1212.4984}{{\ttfamily
  arXiv:1212.4984 [hep-th]}}.

\bibitem{Sumitomo:2013vla}
Y.~Sumitomo, S.~Tye, and S.~S. Wong, ``{Statistical Distribution of the Vacuum
  Energy Density in Racetrack K{\"a}hler Uplift Models in String Theory},''
  \href{http://dx.doi.org/10.1007/JHEP07(2013)052}{{\em JHEP} {\bfseries 07}
  (2013) 052}, \href{http://arxiv.org/abs/1305.0753}{{\ttfamily arXiv:1305.0753
  [hep-th]}}.

\bibitem{Tye:2016jzi}
S.~H.~H. Tye and S.~S.~C. Wong, ``{Linking Light Scalar Modes with A Small
  Positive Cosmological Constant in String Theory},''
  \href{http://dx.doi.org/10.1007/JHEP06(2017)094}{{\em JHEP} {\bfseries 06}
  (2017) 094}, \href{http://arxiv.org/abs/1611.05786}{{\ttfamily
  arXiv:1611.05786 [hep-th]}}.

\bibitem{Giddings:2001yu}
S.~B. Giddings, S.~Kachru, and J.~Polchinski, ``{Hierarchies from fluxes in
  string compactifications},''
  \href{http://dx.doi.org/10.1103/PhysRevD.66.106006}{{\em Phys. Rev. D}
  {\bfseries 66} (2002) 106006},
  \href{http://arxiv.org/abs/hep-th/0105097}{{\ttfamily arXiv:hep-th/0105097}}.

\bibitem{polya}
G.~P\'olya, ``{About a task of the probability calculation concerning the
  randomness in the road network},'' {\em Math. Annals} {\bfseries 84} (1921)
  149.

\bibitem{Svrcek:2006yi}
P.~Svrcek and E.~Witten, ``{Axions In String Theory},''
  \href{http://dx.doi.org/10.1088/1126-6708/2006/06/051}{{\em JHEP} {\bfseries
  06} (2006) 051}, \href{http://arxiv.org/abs/hep-th/0605206}{{\ttfamily
  arXiv:hep-th/0605206}}.

\bibitem{Banks:2003sx}
T.~Banks, M.~Dine, P.~J. Fox, and E.~Gorbatov, ``{On the possibility of large
  axion decay constants},''
  \href{http://dx.doi.org/10.1088/1475-7516/2003/06/001}{{\em JCAP} {\bfseries
  06} (2003) 001}, \href{http://arxiv.org/abs/hep-th/0303252}{{\ttfamily
  arXiv:hep-th/0303252}}.

\bibitem{grinstein}
G.~Grinstein, D.-H. Lee, and S.~Sachdev, ``{Conservation laws, anisotropy, and
  ``self-organized criticality" in noisy nonequilibrium systems},'' {\em Phys.
  Rev. Lett.} {\bfseries 64} (1990) 1927.

\bibitem{Odor:2002hk}
G.~Odor, ``{Phase transition universality classes of classical, nonequilibrium
  systems},'' \href{http://dx.doi.org/10.1103/RevModPhys.76.663}{{\em Rev. Mod.
  Phys.} {\bfseries 76} (2004) 663},
  \href{http://arxiv.org/abs/cond-mat/0205644}{{\ttfamily
  arXiv:cond-mat/0205644}}.

\bibitem{Doye_2002}
J.~P.~K. Doye, ``{Network Topology of a Potential Energy Landscape: A Static
  Scale-Free Network},'' {\em Phys. Rev. Lett.} {\bfseries 88} (2002) 238701.

\bibitem{Rao_Caflisch_2004}
F.~Rao and A.~Caflisch, ``{The protein folding network},'' {\em J. Mol. Biol.}
  {\bfseries 342} (2009) 299--306,
  \href{http://arxiv.org/abs/q-bio/0403034}{{\ttfamily arXiv:q-bio/0403034
  [q-bio.BM]}}.

\bibitem{doi:10.1063/1.4990866}
T.~Weng, J.~Zhang, M.~Small, J.~Yang, F.~H. Bijarbooneh, and P.~Hui,
  ``Multitarget search on complex networks: A logarithmic growth of global mean
  random cover time,'' \href{http://dx.doi.org/10.1063/1.4990866}{{\em Chaos:
  An Interdisciplinary Journal of Nonlinear Science} {\bfseries 27} no.~9,
  (2017) 093103}. \url{https://doi.org/10.1063/1.4990866}.

\bibitem{cover_time}
D.~J. Aldous, ``On the time taken by random walks on finite groups to visit
  every state,'' {\em {Zeitschrift f\"ur Wahrscheinlichkeitstheorie und
  Verwandte Gebiete}} {\bfseries 63} (1983) 361--374.

\end{thebibliography}\endgroup
\end{document}